\documentclass{emulateapj}
\usepackage{amsmath}
\usepackage{natbib}
\usepackage{multirow}
\usepackage{graphicx}
\usepackage{epstopdf}
\newcommand{\gps}{\ensuremath{g_{\rm P1}}}
\newcommand{\rps}{\ensuremath{r_{\rm P1}}}
\newcommand{\ips}{\ensuremath{i_{\rm P1}}}
\newcommand{\zps}{\ensuremath{z_{\rm P1}}}
\newcommand{\yps}{\ensuremath{y_{\rm P1}}}

\newcommand{\PS}{\protect \hbox {Pan-STARRS1}}
\newcommand{\degree}{\ensuremath{^\circ}}
\newcommand{\dfplot}[1]{\plotone{#1}}

\newcommand{\m}{\ensuremath{\vec{m}}}

\newcommand{\mum}{\ensuremath{\mu}m}

\newcommand{\Ep}{\ensuremath{E^\prime}}
\newcommand{\RpV}{\ensuremath{R^\prime(V)}}
\newcommand{\R}{\ensuremath{\vec{R_0}}}
\newcommand{\dRdx}{\ensuremath{d\vec{R}/dx}}

\shorttitle{The Optical-Infrared Extinction Curve and its Variation in the Milky Way}
\shortauthors{E. F. Schlafly et al.}
\begin{document}
\title{The Optical-Infrared Extinction Curve and its Variation in the Milky Way}
\author{
E. F. Schlafly,\altaffilmark{1,2,3}
A. M. Meisner,\altaffilmark{4,2}
A. M. Stutz,\altaffilmark{1}
J. Kainulainen,\altaffilmark{1}
J. E. G. Peek,\altaffilmark{5}
K. Tchernyshyov,\altaffilmark{6}
H.-W. Rix,\altaffilmark{1}
D. P. Finkbeiner,\altaffilmark{7,8}
K. R. Covey,\altaffilmark{9}
G. M. Green,\altaffilmark{7}
E. F. Bell,\altaffilmark{10}
W. S. Burgett,\altaffilmark{11}
K. C. Chambers,\altaffilmark{12}
P. W. Draper,\altaffilmark{13}
H. Flewelling,\altaffilmark{12}
K. W. Hodapp,\altaffilmark{12}
N. Kaiser,\altaffilmark{12}
E. A. Magnier,\altaffilmark{12}
N. F. Martin,\altaffilmark{14,1}
N. Metcalfe,\altaffilmark{13}
R. J. Wainscoat,\altaffilmark{12}
C. Waters\altaffilmark{12}
}

\altaffiltext{1}{Max Planck Institute for Astronomy, K\"{o}nigstuhl 17, D-69117 Heidelberg, Germany}
\altaffiltext{2}{Lawrence Berkeley National Laboratory, One Cyclotron Road, Berkeley, CA 94720, USA}
\altaffiltext{3}{Hubble Fellow}
\altaffiltext{4}{Berkeley Center for Cosmological Physics, Berkeley, CA 94720, USA}
\altaffiltext{5}{Space Telescope Science Institute, 3700 San Martin Dr, Baltimore, MD 21218, USA}
\altaffiltext{6}{Department of Physics and Astronomy, The Johns Hopkins University, 3400 North Charles Street, Baltimore, MD 21218, USA}
\altaffiltext{7}{Harvard-Smithsonian Center for Astrophysics, 60 Garden Street, Cambridge, MA 02138, USA}
\altaffiltext{8}{Department of Physics, Harvard University, 17 Oxford Street, Cambridge MA 02138, USA}
\altaffiltext{9}{Department of Physics and Astronomy, Western Washington University, 516 High St., Bellingham, WA 98225, USA}
\altaffiltext{10}{Department of Astronomy, University of Michigan, 500 Church St., Ann Arbor, MI 48109, USA}
\altaffiltext{11}{GMTO Corporation, 251 S. Lake Ave., Suite 300, Pasadena, CA  91101, USA}
\altaffiltext{12}{Institute for Astronomy, University of Hawaii, 2680 Woodlawn Drive, Honolulu HI 96822, USA}
\altaffiltext{13}{Department of Physics, Durham University, South Road, Durham DH1 3LE, UK} 
\altaffiltext{14}{Observatoire Astronomique de Strasbourg, CNRS, UMR 7550, 11 rue de l'Universit\'{e}, F-67000 Strasbourg, France}

\begin{abstract}
The dust extinction curve is a critical component of many observational programs and an important diagnostic of the physics of the interstellar medium.  Here we present new measurements of the dust extinction curve and its variation towards tens of thousands of stars, a hundred-fold larger sample than in existing detailed studies.  We use data from the APOGEE spectroscopic survey in combination with ten-band photometry from Pan-STARRS1, 2MASS, and WISE.  We find that the extinction curve in the optical through infrared is well characterized by a one-parameter family of curves described by $R(V)$.  The extinction curve is more uniform than suggested in past works, with $\sigma(R(V)) = 0.18$, and with less than one percent of sight lines having $R(V) > 4$.  Our data and analysis have revealed two new aspects of Galactic extinction: first, we find significant, wide-area variations in $R(V)$ throughout the Galactic plane.  These variations are on scales much larger than individual molecular clouds, indicating that $R(V)$ variations must trace much more than just grain growth in dense molecular environments.  Indeed, we find no correlation between $R(V)$ and dust column density up to $E(B-V) \approx 2$.  Second, we discover a strong relationship between $R(V)$ and the far-infrared dust emissivity.
\end{abstract}


\section{Introduction}
\label{sec:intro}

Dust is composed of small, solid grains of material.  These grains are made of heavy elements formed from nuclear fusion in stars, blown into the interstellar medium (ISM) by stellar winds and explosions.  The dust grains scatter and absorb light.  Owing to the small size of the grains, dust preferentially scatters and absorbs blue light relative to red light in the optical through infrared.  The resulting extinction as a function of wavelength is called the dust extinction curve \citep{Draine:2003}.

Effective parameterizations of the dust extinction curve in the ultraviolet (UV) and optical were developed by \citet{Fitzpatrick:1986, Fitzpatrick:1988}.  The work of \citet[CCM]{Cardelli:1989} showed that much of the variation could be described by a single parameter, $R(V) = A(V) / E(B-V)$, the total-to-selective extinction ratio, though especially in the UV this description is far from complete.

The shape of the extinction curve is a valuable diagnostic of the properties of the dust.  Variation in $R(V)$ is sometimes attributed to variation in the size distribution of dust grains; dust with high $R(V)$ has a relatively gray, flat extinction curve in the optical, suggesting an abundance of large grains relative to small.  Variation in $R(V)$ may also be related to the formation of ice on dust grains or grain aggregation in dense environments ($E(B-V) > 1$) \citep{Whittet:1988, Ysard:2013}.  Current observational evidence suggests a relationship between $R(V)$ and $E(B-V)$, though the strength of the correlation is not clear \citep{Fitzpatrick:2007, Foster:2013}.  Alternatively, variation in $R(V)$ may be driven by grain processing by UV photons or grain composition and chemistry \citep{Jones:2013, Mulas:2013}.  

Many studies of the extinction curve have focused on relatively small samples of O and B stars or on particular molecular clouds.  For example, the best atlases of Milky Way extinction curves are those of \citet{Valencic:2004} and \citet{Fitzpatrick:2007}, which have samples of a few hundred O and B stars.  These stars are practical targets for studies of the extinction curve in the UV, owing to their intrinsic UV brightness.  However, the small number of these stars and the atypical environments they inhabit make their use to characterize the variability of the extinction curve problematic.  Additionally, the sparseness of appropriate bright O and B targets complicates morphological association of the observed extinction curve variations with known structures in the interstellar medium, inhibiting efforts to uncover the underlying physical processes at work.

Many studies of the extinction curve have focused on only a small range of wavelengths for practical reasons, for example, focusing either on only the optical \citep[e.g.][]{Schlafly:2010} or only the near-infrared \citep[e.g.][]{Wang:2014} extinction curve.  This means that individual parts of the extinction curve are known much better than the connections between these parts.

Perhaps unsurprisingly, then, current descriptions of the extinction curve are in substantial tension.  For example, the CCM extinction curve contains no variation in the infrared, while the extinction curve of \citet{Fitzpatrick:2009} is more variable in the near-infrared than in the red-optical.  Different authors have found both ``universal'' near-infrared extinction \citep[e.g.][]{Wang:2014} and variable infrared extinction \citep[e.g.][]{Zasowski:2009}.  Meanwhile optical studies essentially always find extinction curve variability, though often with the caveat that ``most'' extinction curves are compatible with an ``average'' Milky Way extinction curve \citep[e.g.][]{Krelowski:2012}.  That said, the detailed shape of the optical variation differs significantly between, for example, the extinction curves of CCM, \citet{Fitzpatrick:2007}, and \citet{MaizApellaniz:2014}.

Here we present a far more comprehensive, multiwavelength study of the Galactic dust extinction curve and its variation, combining data from spectroscopic and photometric surveys.  The APOGEE survey has spectroscopically observed about 150,000 stars in the Galactic midplane, obtaining accurate temperatures, metallicities, and gravities \citep{Majewski:2015}.  The Pan-STARRS1 survey has photometrically observed the entire sky north of declination $-30\degree$, providing optical photometry for essentially all of these stars.  Infrared photometry from 2MASS \citep{Skrutskie:2006} and WISE \citep{Wright:2010} complement the optical photometry, providing coverage over a factor of ten in wavelength.  The APOGEE targets span Galactic longitudes from roughly $0\degree < l < 240\degree$, with typical distances between 1 and 5 kpc.  The broad coverage of the Galactic plane provides an excellent test for studying the extinction curve and its variation throughout the Milky Way.

We study extinction via a generalization of the ``pair method,'' where unextinguished ``standard'' stars are compared with extinguished stars of the same spectral type to assess the extinction to those stars.  This technique was pioneered by \citet{Trumpler:1930}, but large spectroscopic surveys have recently allowed it to be applied to hundreds of thousands of stars \citep[e.g.][]{Schlafly:2011, Yuan:2013}.  It is impossible to apply the pair method directly to the APOGEE targets, because APOGEE has observed very few unextinguished stars; almost all of the sight lines observed are heavily extinguished.  Moreover, the few unextinguished stars in the survey are typically low metallicity halo stars, while the extinguished stars are metal-rich disk stars, complicating the comparison of unextinguished standards with extinguished targets.  We circumvent this difficulty by focusing on the {\it shape} of the extinction curve, leaving the extinction to any individual star relatively unconstrained: we compare stars of the same stellar types, but behind different amounts of extinction to determine the extinction curve.

An alternative approach to solving this issue using synthetic stellar spectra has been extensively explored in the literature \citep{Fitzpatrick:2005, Fitzpatrick:2007, Fitzpatrick:2009, Schultheis:2014, Schultheis:2015, Clayton:2015}.  In this approach, the problem of finding unreddened standard stars is avoided by using theoretical stellar spectra instead of observed spectra.  We eschew this approach here in order to limit our exposure to any systematic errors in the theoretical spectra.  In past work we have found such errors to be of the order of a few percent \citep{Schlafly:2011}, which is a few times larger than typical observational uncertainties.

This paper is divided into several sections.  First, in \textsection\ref{sec:data}, we discuss the observational data: the APOGEE spectroscopy, and \PS, 2MASS, and WISE photometry.  In \textsection\ref{sec:method}, we discuss our technique for studying the extinction curve and its variation with this data.  In \textsection\ref{sec:results}, we show the results of our analysis.  In \textsection\ref{sec:discussion}, we compare these results with the literature and discuss the consequences for our understanding of dust in the Milky Way.  Finally, in \textsection\ref{sec:conclusion}, we conclude.

\section{Data}
\label{sec:data}

\subsection{APOGEE}

The APOGEE survey is a high-resolution ($R = 22500$) spectroscopic, near-infrared (NIR) H-band survey of the sky \citep{Majewski:2015}.  The APOGEE spectrograph \citep{Wilson:2010} is illuminated with 300 fibers from the Sloan Digital Sky Survey (SDSS) 2.5~m telescope \citep{Gunn:2006}.  The APOGEE survey is largely focused on obtaining abundances of distant giant stars in the Galactic disk, and so observes to a high signal-to-noise ratio of 100 per half-resolution element.  The APOGEE spectroscopic pipeline \citep{Nidever:2015} and astrophysical parameter pipeline \citep{GarciaPerez:2015} measure temperatures, metallicities, and gravities for the stars, with typical uncertainties of $<100~\mathrm{K}$, $<0.05~\mathrm{dex}$, and $<0.15~\mathrm{dex}$, respectively.  The main APOGEE survey targets red clump and red giant stars based on their dereddened NIR color \citep{Zasowski:2013}.  The APOGEE spectroscopic parameters are determined from continuum-normalized spectra, and are therefore insensitive to the extinction of the source.  This work uses data from SDSS-III Data Release 12 \citep{Eisenstein:2011, Alam:2015, Holtzman:2015}, containing spectra of more than 150,000 stars.

\subsection{WISE}
The {\it Wide-field Infrared Survey Explorer} is a NASA infrared space telescope that has surveyed the entire sky at 3.4 (W1), 4.6 (W2), 12 (W3), and 22~\mum\ (W4) wavelengths \citep{Wright:2010}.  We use W1 and W2 photometry from the AllWISE data release \citep{Cutri:2013}, which contains data from two complete sky coverage epochs and has identified more than 700 million objects.  Approximately one third of all WISE detections of APOGEE targets have flags indicating problematic conditions (\texttt{cc\_flags}), usually due to being in the halo of a nearby bright star or landing on a diffraction spike.  We conservatively assign infinite uncertainties to the WISE photometry of these sources.

\subsection{2MASS}
The Two Micron All-Sky Survey (2MASS) was a near-infrared survey of the entire sky in the $J$ (1.25 \mum), $H$ (1.65 \mum), and $K$ (2.17 \mum) bands, undertaken from 1997--2001 \citep{Skrutskie:2006}.  The 2MASS survey contains observations of more than 300 million objects, including all objects targeted by APOGEE.

\subsection{Pan-STARRS1}
\label{subsec:ps1}
We obtain optical photometry of APOGEE targets from the \PS\ survey.  The \PS\ observations are made on a 1.8~m telescope on Haleakala \citep{PS1_optics}.  The telescope focal plane is outfitted with the 1.4 billion pixel GPC1 camera \citep{PS1_optics, PS1_GPCA, PS1_GPCB}, which covers the 3\degree\ field of view of the telescope.  Observations are performed in five broad passbands, covering about 400~nm to 1~\mum\ \citep{PS_lasercal}.  The effective wavelengths of the filters are roughly 480, 620, 750, 870, and 960~nm, for the \gps, \rps, \ips, \zps, and \yps\ filters, respectively.  The data are automatically processed by the \PS\ Image Processing Pipeline \citep{PS1_IPP}, which analyzes images to deliver photometry, astrometry, and morphology \citep{PS1_photometry, PS1_astrometry, Magnier:2013}.  The photometric calibration of the survey, both relative and absolute, is accurate to better than 1\% \citep{JTphoto, Schlafly:2012}.

Many stars in APOGEE are too bright to have reliable photometric magnitudes in PS1.  In our analysis, we mark the uncertainties of any measurements brighter than 14.0, 14.4, 14.4, 13.8, and 13.3 magnitudes in the \gps, \rps, \ips, \zps, and \yps\ bands as infinitely large to reflect the increased systematic uncertainties brightward of these limits.

\subsection{Target Selection}
\label{subsec:targetselection}

This work uses only 37,000 stars of the full 150,000 stars in APOGEE DR12.  The most significant restriction we adopt is to use only stars for which we estimate the PS1 \yps\ magnitude is fainter than 13th magnitude, estimated from the 2MASS colors and the Rayleigh-Jeans Color Excess estimated extinction \citep{Zasowski:2013}.  Stars not passing this cut will be saturated in many of the optical bands, limiting their use in characterizing the extinction curve.  We also exclude any stars targeted as part of APOGEE ``ancillary'' programs, to include only stars from the main survey.  Ancillary program stars may have selection criteria making them unsuitable for reddening studies.  We further exclude any stars where accurate gravities could not be determined, or where the APOGEE flags indicate that stars had an abnormally high $\chi^2$ or rotation rate.  Lastly, we exclude any stars whose PS1 or WISE position is more than $0.5^{\prime\prime}$  separated from its 2MASS position.  These cuts reduce the more than 150,000 stars in APOGEE DR12 to under 40,000 stars.  Most of the stars are removed by the brightness cut (150,000 $\rightarrow$ 72,000).  The requirement of a reliable gravity reduces the number further to 47,000, and the elimination of ancillary targets, stars with problematic flags, and astrometric separation reduces the number to 37,000, our full data set.

\begin{figure*}[htb]
\dfplot{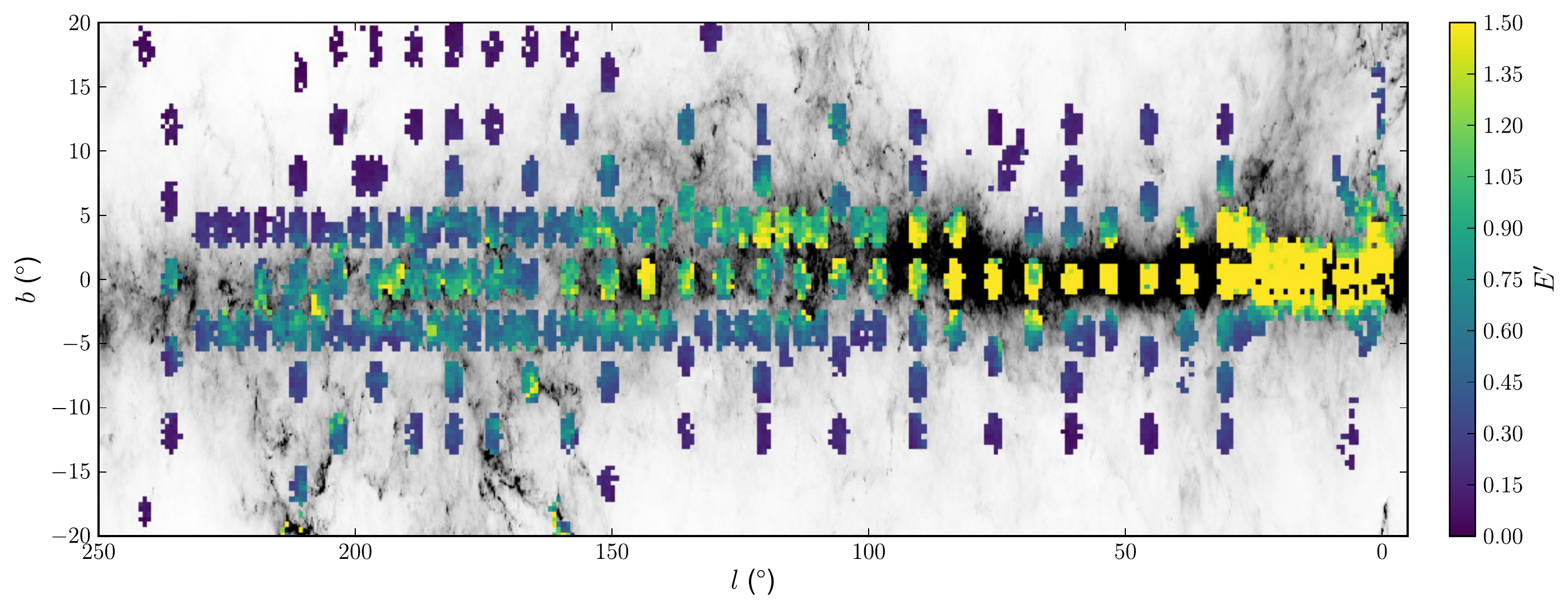}
\caption[\Ep\ of APOGEE Targets]{
\label{fig:ebvmap}
\Ep\ (roughly $E(B-V)$) to APOGEE targets.  The best fit \Ep\ is shown as colored points.  The background grayscale shows the Planck $\tau_{353}$-based extinction map \citep{Planck:2014} for context, and ranges from 0--2.5~mag \Ep.  Unsurprisingly, the most reddened APOGEE targets are at low latitudes in the inner Galaxy.
}
\end{figure*}

\begin{figure}[htb]
\dfplot{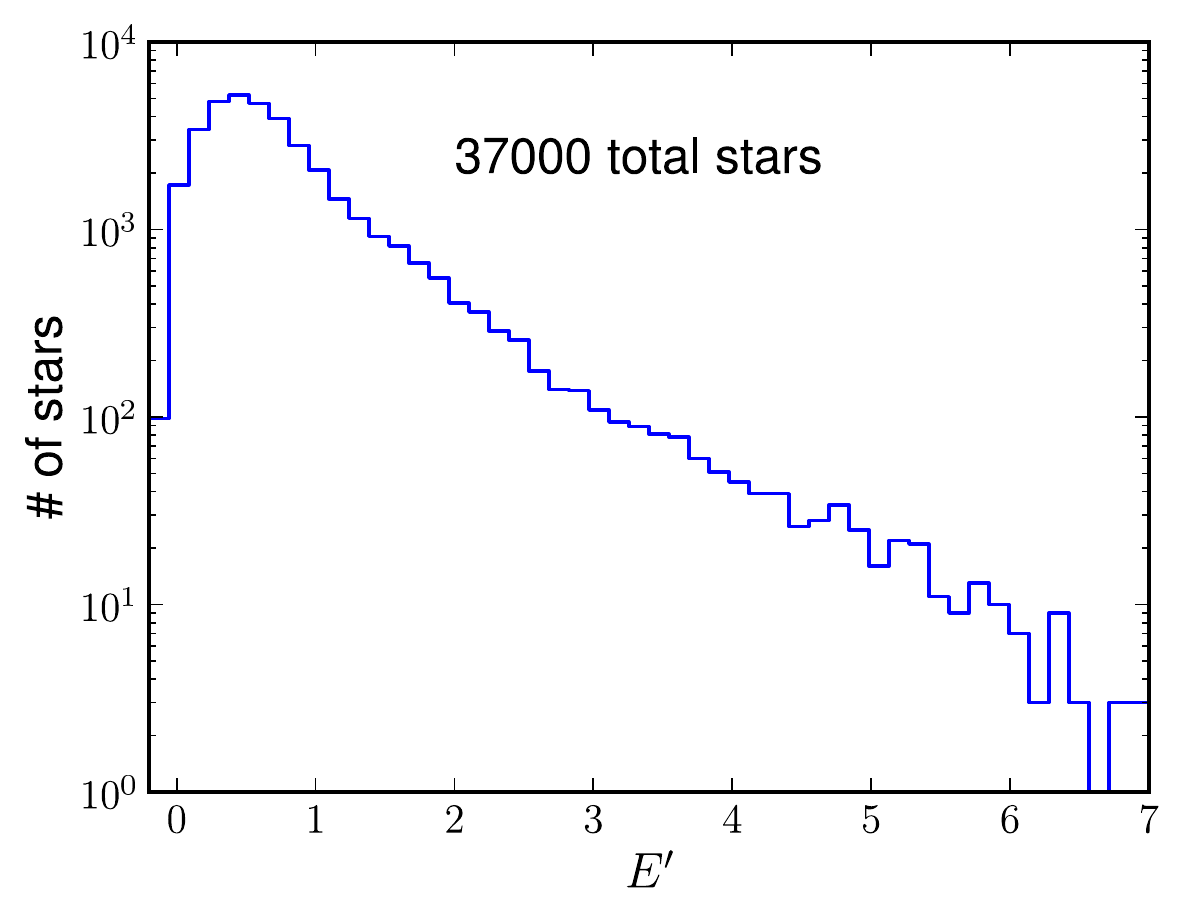}
\caption[\Ep\ histogram]{
\label{fig:ebvhist}
Histogram of \Ep\ (roughly $E(B-V)$) to APOGEE targets.  The typical \Ep\ is roughly 0.65, though the distribution is broad and ranges from 0 to more than 5 mag.
}
\end{figure}

We show in Figure~\ref{fig:ebvmap} the locations of the stars used in this work.  Points are colored by their estimated reddenings \Ep\ (roughly, $E(B-V)$), as determined in this work.  The background image shows the Planck $\tau_{353}$-based extinction map \citep{Planck:2014} for context.  Figure~\ref{fig:ebvhist} shows a histogram of the reddenings to these stars.  The mean reddening is 0.65~mag with a standard deviation of 0.5, though the distribution extends to $\Ep > 5$ mag.  Not all stars are detected in all photometric bands: for instance, when $\Ep > 2.5$ mag, most stars are not detected in \gps.

\section{Method}
\label{sec:method}

Our basic technique is to model the observed ten optical-infrared magnitudes of the APOGEE stars as a function of their temperature and metallicity, as well as their distance and reddening.  The analysis has two parts.  In the first part, we assume the extinction curve is universal, completely non-varying throughout the Galaxy.  We then determine the fixed shape of the extinction curve and the intrinsic colors of stars as a function of their temperatures and metallicities.  In the second part, we use the previously-determined intrinsic colors to determine the reddening to each star in each band.  With these reddenings in hand, we are able to relax our initial assumption that the extinction curve has a single fixed shape, and study the extinction curve's variation throughout the Galaxy via a kind of principal component analysis.

This two-step approach allows us to initially fit a complicated function describing the intrinsic colors of stars as a function of temperature and metallicity, while adopting a simple description of the extinction.  After we have intrinsic colors, we assume in the second step that these are known.  This makes it possible to fit a more complicated model for the effect of reddening on the observed colors, where we allow the shape of the extinction curve to vary.  

Our separation of the problem into two steps is not strictly optimal.  To illustrate this point with an extreme example, consider a case where every star of a particular temperature and metallicity were behind an identical column of dust with a ``peculiar'' extinction curve.  Then our initial fit would obtain incorrect intrinsic colors for stars of that temperature and metallicity.  Our extinction curve variability fit would then find completely ordinary extinction for these stars, since the ``peculiar'' extinction would have been absorbed into the intrinsic colors and removed.  

Fortunately, this situation is contrived: throughout the Galaxy we find a wide range of stellar types, and the dust column density varies on much smaller scales than the large scales over which the Galaxy's stellar population varies.  Moreover, by comparing our intrinsic colors with predictions from synthetic spectra, we verify that the intrinsic colors we obtain are reasonable.

\subsection{Initial Fit}
\label{subsec:initialfit}

In our initial fit, we assume that the extinction curve is universal, described by a single, fixed reddening vector \R\ that is the same for all stars we consider.  We then model the observed photometry $\m$ as a function of the temperature and metallicity of the star, as well as the star's distance and reddening.  To be specific, we model the stars' photometry as
\begin{equation}
\label{eq:modelmag}
  \m_i^m = \vec{f}(T_i, \mathrm{[Fe/H]}_i) + \mu_i + \R \Ep_i \, ,
\end{equation}
where $i$ indexes over stars, and each vector $\m_i^m$ contains the $i$th star's photometry in the 10 photometric bands.  In this equation, $T_i$ and $\mathrm{[Fe/H]}_i$ are the temperature and metallicity of the $i$th star, as determined by APOGEE.  The parameters $\Ep_i$ and $\mu_i$ are the model distance modulus and dust column to the $i$th star.  The function $\vec{f}$ is an analytic function giving the intrinsic colors as a function of temperature and metallicity, with each element of $\vec{f}$ corresponding to a particular band.  We parameterize $\vec{f}$ in each band as a fourth-order polynomial in $T$ and $\mathrm{[Fe/H]}$, including cross terms, and additionally including terms proportional to $T^5$ and $T^6$.  This means that 17 parameters are needed to describe $\vec{f}$ in each band, for a total of 170 free parameters.  Finally, \R\ is the fixed reddening vector giving the relative amount of extinction in each band, described by 10 parameters.

To completely determine the model, then, the following global parameters must be specified:
\begin{itemize}
\item the intrinsic color function $\vec{f}$, as determined by the 17 coefficients of the polynomial in each of the 10 bands, and
\item the reddening vector $\R$, as determined by the 10 elements giving the relative extinctions in each band.
\end{itemize}
In addition to this, the following two parameters must be specified for each of the 37000 stars:
\begin{itemize}
\item the distance modulus $\mu_i$, and
\item the dust column $\Ep_i$.
\end{itemize}
The complete model then predicts the magnitudes of each star in each of the 10 bands according to Equation~\ref{eq:modelmag}.  

For a given choice of these model parameters, we evaluate $\chi$, the difference between the data and the model, scaled by the uncertainty
\begin{equation}
\label{eq:chi}
  \chi_{i,j} = \frac{m_{i,j} - m^m_{i,j}}{\sigma_{i,j}} \, ,
\end{equation}
where $i$ indexes over the different stars, $j$ indexes over the different photometric bands, $\m$ is the observed photometry, $\vec{\sigma}$ is its uncertainty, and $\m^m$ is given in Equation~\ref{eq:modelmag}.

We determine the best fit model parameters by minimizing the sum of $\chi^2$.  In practice, Equation~\ref{eq:chi} is vulnerable to outliers, and so we instead minimize
\begin{equation}
\label{eq:chi2damp}
  \sum_{i,j} \hat{\chi}_{i,j}^2 = \sum_{i,j} \left(\frac{\chi_{i,j}}{1+\sqrt{|\chi_{i,j}|/5}}\right)^2 \, ,
\end{equation}
smoothly switching from minimizing $\chi^2$ to minimizing $|\chi|$ at $\chi = 5$, philosophically similar to clipping at $5\sigma$.

The model requires a total of 74,180 parameters to be fit (ignoring, for a moment, 36 parameters which are subject to perfect degeneracies and are fixed; see \textsection\ref{subsubsec:degeneracies}).  This is a substantial number, but they are constrained by roughly 300,000 photometric measurements, and so the model is well constrained.  

We minimize Equation~\ref{eq:chi2damp} to determine the best fit parameters using an alternating-least-squares algorithm.  We first fix the global model parameters describing $\vec{f}$ and \R, and solve for the parameters \Ep\ and $\mu$ for each star, one by one.  We then fix these values of \Ep\ and $\mu$, and solve for the global model parameters $\vec{f}$ and \R.  With the improved global parameters, we then solve for \Ep\ and $\mu$ again, iterating back and forth between solving for the per-star and global parameters until converged.

The fit has two important results.  First, it gives us $\vec{f}(T, \mathrm{[Fe/H]})$, the intrinsic colors of the stars as a function of their APOGEE spectroscopic parameters.  With these in hand, the observed reddenings are trivially computed as $\m - \vec{f}(T, \mathrm{[Fe/H]})$, modulo a gray component of extinction and the distances to the sources.  Second, the fit gives the mean extinction curve over the APOGEE footprint, as encoded in \R.

\subsubsection{Degeneracies}
\label{subsubsec:degeneracies}

This model is subject to a number of perfect degeneracies which we eliminate by fixing certain parameters.  It is instructive to consider these degeneracies to see what signals we are sensitive to.

First, we are insensitive to any gray component of extinction ($\R \rightarrow \R + C$).  Any gray component of extinction can be absorbed by appropriately adjusting $\mu$ for each star $i$ ($\mu_i \rightarrow \mu_i - C\Ep_i$).  Likewise, we are insensitive to any change in the overall brightness of the stars ($\vec{f} \rightarrow \vec{f} + C$), since this can likewise be absorbed into $\mu$.  To address these degeneracies, we fix one component of \R\ and fix $\vec{f}$ to be 0 in the $K$ band (18 parameters in total).

Second, we are insensitive to any change in the normalization of the extinction curve ($\R \rightarrow C\R)$.  Any change in the normalization of the extinction curve can be compensated for by rescaling the reddenings of each star ($\Ep_i \rightarrow \Ep_i/C$).  To remove this degeneracy, we fix a second component of \R\ (1 parameter).

Finally, we are insensitive to any change of $\vec{f}$ along the reddening vector ($\vec{f}(T, \mathrm{[Fe/H]}) \rightarrow \vec{f} + \R g(T, \mathrm{[Fe/H]})$).  For such changes, we could modify the extinction $\Ep_i \rightarrow \Ep_i-g(T_i, \mathrm{[Fe/H]}_i)$ for each star to cancel the effect.  Accordingly, we can choose to set any color of choice in $\vec{f}$ to theoretical expectations.  We choose to set the \yps\ component of $\vec{f}$ to a specific functional form picked to match expectations from synthetic spectra and the few low reddening stars in APOGEE; see \textsection\ref{subsec:intcolors} for further details.  This fixes 17 free parameters.  We note that if we had an adequate set of unreddened standard stars in APOGEE, we could avoid relying on theoretical expectations to fix these parameters.

These degeneracies mean that although $\vec{f}$ is intended to represent stars' absolute magnitudes, it does not: we have no access to absolute magnitudes since we have no distances.  Instead, it ultimately encodes only the intrinsic colors of stars as a function of their stellar parameters.  Likewise, any additive offset to the reddening vector (that is, a gray component) is not measurable by our technique.  This also means that our distance moduli $\mu$ for each star are combinations of the true distance moduli, errors in the gray component of our reddening vector, and errors in our absolute magnitudes.  We do not use these ``distance moduli'' further in this work, and look forward to parallax measurements from Gaia, which will lift this degeneracy.

\subsubsection{Limitations}
\label{subsubsec:limitations}

Our model is correct in the limit that a star's intrinsic colors are polynomial functions of its temperature and metallicity, and that reddening in broad photometric bands can be described by a single, universal vector.  The former assumption is expected to be valid up to the accuracy of the photometry in this work ($\sim 1\%$), though in principle a star's gravity, blended companions, rotation, and detailed abundances may have a small effect on the star's photometry.  

The latter assumption---that reddening can be described by a single vector---is only true when the photometric bandpass is narrow and the extinction curve is universal.  Neither condition applies.  To mitigate the first problem, we could apply the results of \citet{Sale:2015}.  In this case, we could model the effect of reddening on magnitudes by
\begin{equation}
   \m^m = \vec{f}(T, \mathrm{[Fe/H]}) + \mu + \R \Ep + \vec{g}(T, \mathrm{[Fe/H]}, \log g, \Ep) \, ,
\end{equation}
where $\vec{g}(T, \mathrm{[Fe/H]}, \log g, \Ep)$ is a function describing the effect of reddening in the different bands for stars of different temperatures.  The function $\vec{g}$ can be fixed using existing extinction curves and synthetic stellar spectra as in \citet{Sale:2015}.  In this parameterization, the reddening vector \R\ would be a small perturbation to an existing extinction curve.  We explored initially choosing $\vec{g}$ to reproduce the extinction curve of \citet{Fitzpatrick:2009}, assuming intrinsic stellar spectra given by the MARCS model grid \citep{Gustafsson:2008}.  However, when applying this model, we found no significant improvement in $\chi^2$.  The primary effect of the more principled treatment was to scale the reddenings of highly reddened stars a few percent higher, but we are largely unconcerned here with the accuracy of the inferred monochromatic extinctions to individual stars, and so have instead applied the simpler treatment where the effect of reddening is linear and independent of the source spectrum.

We have chosen to describe $\vec{f}$ as a fourth-order polynomial, plus terms proportional to $T^5$ and $T^6$.  This places a limit on how well we can reproduce the intrinsic colors of stars.   The choice of polynomial was driven by the desire to reproduce the intrinsic colors to 1\% accuracy.  We experimented with a number of different parameterizations of the intrinsic colors and examined the residuals for trends in temperature and metallicity to determine whether or not we had allowed $\vec{f}$ sufficient freedom to describe the data.  We note that the basic extinction curve results are largely insensitive to the order of the polynomial: the higher-order terms in the polynomial were motivated by reproducing the sharp curve of the intrinsic colors in the $\gps$ band, but these cool stars compose only about 10\% of the whole sample.  The need for a higher order polynomial in $T$ than in $\mathrm{[Fe/H]}$ is due to to the greater dependence of the broadband photometry on temperature than on $\mathrm{[Fe/H]}$, especially in the optical bands.

\subsection{Extinction Curve Variation Fit}
\label{subsec:pca}

We are additionally interested in the variation of the extinction curve in the Galaxy, though we have assumed it to be universal in the previous step.  Since the intial fit (\textsection\ref{subsec:initialfit}) in concert with the APOGEE spectroscopic parameters gives us the intrinsic colors of each star, we can easily compute the reddenings of each star.  Were the extinction curve universal, the observed reddenings would fall along a single line given by the reddening vector, with small dispersion due to photometric and spectroscopic uncertainties, and a small additional scatter owing to the different temperatures of the stars.  Insofar as the extinction curve is in fact a single parameter family---characterized, for instance, by $R(V)$, as in \citet{Cardelli:1989}---then this line will broaden into a 2D surface.  Additional parameters will broaden the surface into higher dimensional manifolds.  In the limit that departures from a universal extinction curve are small, we can linearize the manifolds into linear subspaces.  Accordingly, we can study the variation of the extinction curve by finding the low-dimensional subspaces that best explain the measured reddenings.

We find the best fit mean extinction curve and multi-parameter families of extinction curves by finding low-dimensional subspaces of the ten dimensional space of observed reddenings that best explain the data.  This procedure is essentially a weighted principal component analysis (PCA), with separate weights ($\sigma^{-2}$) for each observation \citep{Jolliffe:2002}.  We find these low-dimensional subspaces via the Heteroscedastic Matrix Factorization technique of \citet{Tsalmantza:2012} \citep[see also][]{Gabriel:1979, Roweis:1998, Tamuz:2005}.  This technique, in contrast to classical PCA, appropriately accounts for the heteroscedastic uncertainty in the observations.  In analogy with PCA, we call the vectors in these subspaces principal components, and order them according to the first subspace in which they appear.

The resulting principal components and the amount of variation in the data along each principal component describe the way in which the extinction curve varies and the significance of that variation.

We note that it is important to perform a heteroscedastic analysis rather than a classical PCA.  Only roughly 30\% of the stars we consider have photometry in all 10 bands, largely due to contaminated WISE photometry in the inner Galactic plane, but also due to saturatation of unreddened stars in the PS1 bands.  Moreover, the uncertainties in the PS1 bands are roughly half those in the infrared bands, leading to a significantly different weighting of the variabilities as compared with classical PCA.

To carry out the analysis quickly, we do repeated $\chi^2$ minimizations to take advantage of the bilinear nature of the problem \citep[see][]{Roweis:1998, Tsalmantza:2012}.  However, $\chi^2$ minimization is vulnerable to outliers, so we again replace $\chi$ by a more robust version according to Equation~\ref{eq:chi}.  To keep the uncertainties diagonal, we perform the analysis in the 10 photometric bands.  However, any gray component of extinction cannot be constrained by our technique.  To address this, we force one vector in the low-dimensional subspace to be the ``gray'' reddening vector; the vectors in the reddening subspace of interest are constrained to be orthogonal to this vector.

We neglect the uncertainty in the model photometry stemming from uncertainty in the APOGEE temperatures and metallicities.  For stars hotter than 4000~K, the uncertainties are close to aligned with the reddening vector and therefore contribute only to a small increase in noise in our estimates for the reddenings to individual stars.  For colder stars, the temperature uncertainties can lead to significant dispersion in colors perpendicular to the reddening vector, which will be identified in this analysis as a reddening signal.  To avoid this, we use only stars with $T > 4000~\mathrm{K}$ to determine the principal components.

\section{Results}
\label{sec:results}

\begin{figure*}[h!tb]
\dfplot{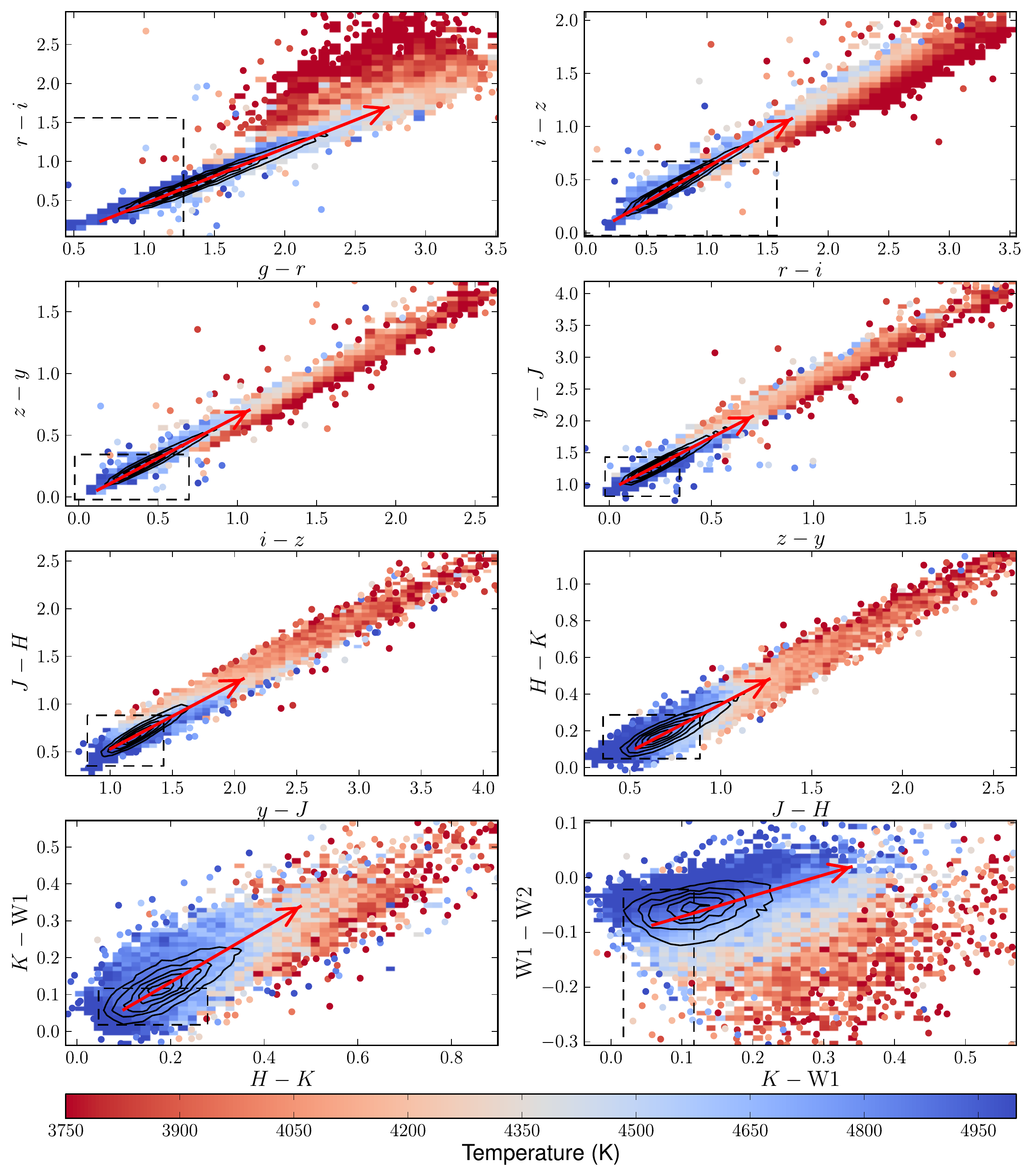}
\caption[Observed Colors]{
\label{fig:data}
The observed colors of APOGEE targets, colored by temperature, in the optical through infrared.  In dense regions of the color-color diagrams, we have replaced individual points with colored bins giving the mean temperature of all points in that bin.  The contours show the number density of points.  The dashed box gives the region where the intrinsic colors lie, shown in more detail in Figure~\ref{fig:obs-model-syncol}.  The red arrow shows the reddening vector we measure.  The observed colors are primarily determined by the stars' reddening (most colors fall right along the reddening vector), though temperature also plays a role.
}
\end{figure*}

\begin{figure*}[h!tb]
\dfplot{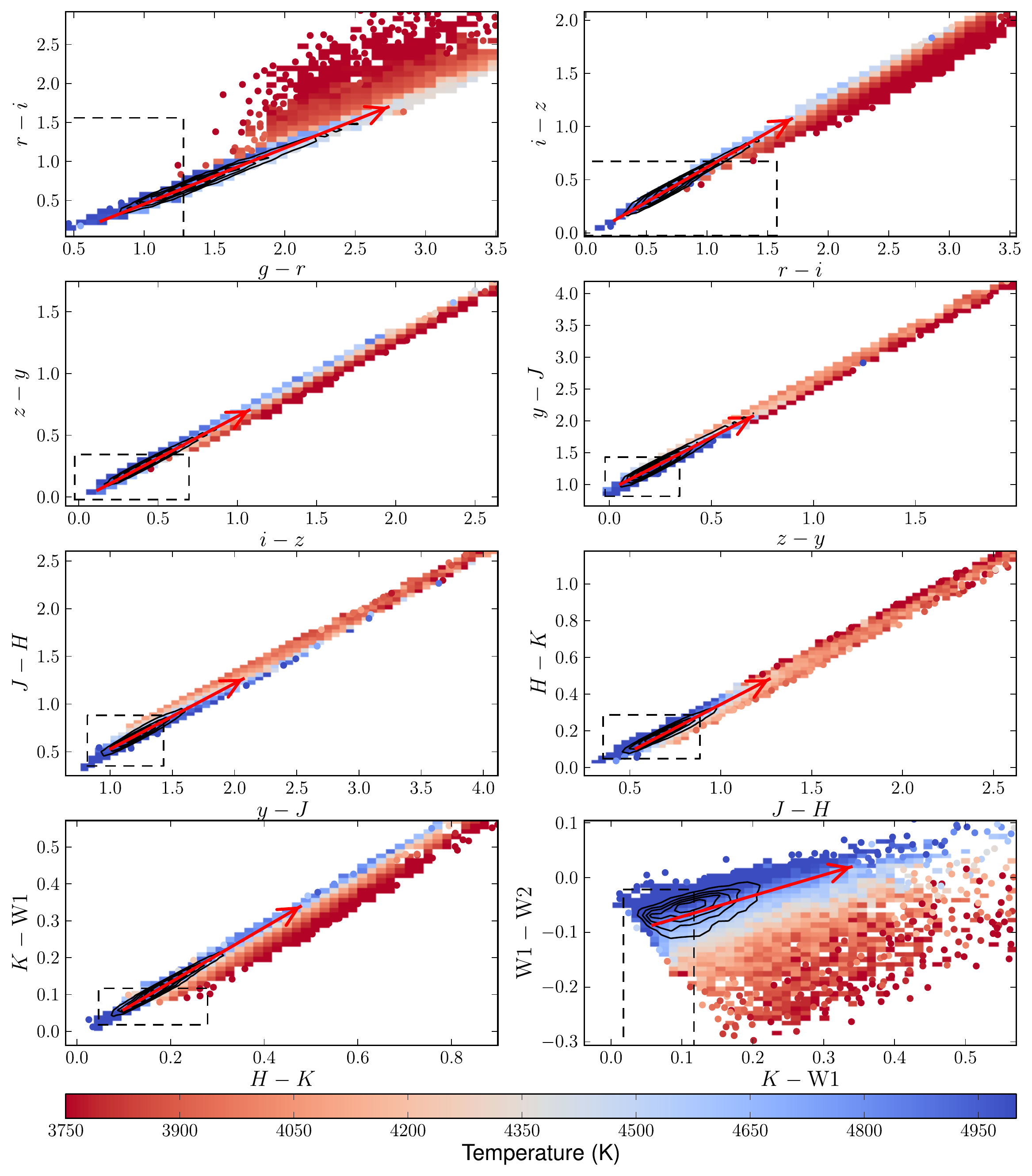}
\caption[Model Colors]{
\label{fig:model}
The best fit model colors of the APOGEE targets.  The figure elements are the same as in Figure~\ref{fig:data}.  These match the observed colors extremely well, modulo the reduced blurring due to noise.  This is especially obvious in the reddest bands, where the observational signal is the smallest, since the WISE colors have much less sensitivity to temperature and reddening than the optical bands.
}
\end{figure*}

\begin{figure*}[h!tb]
\dfplot{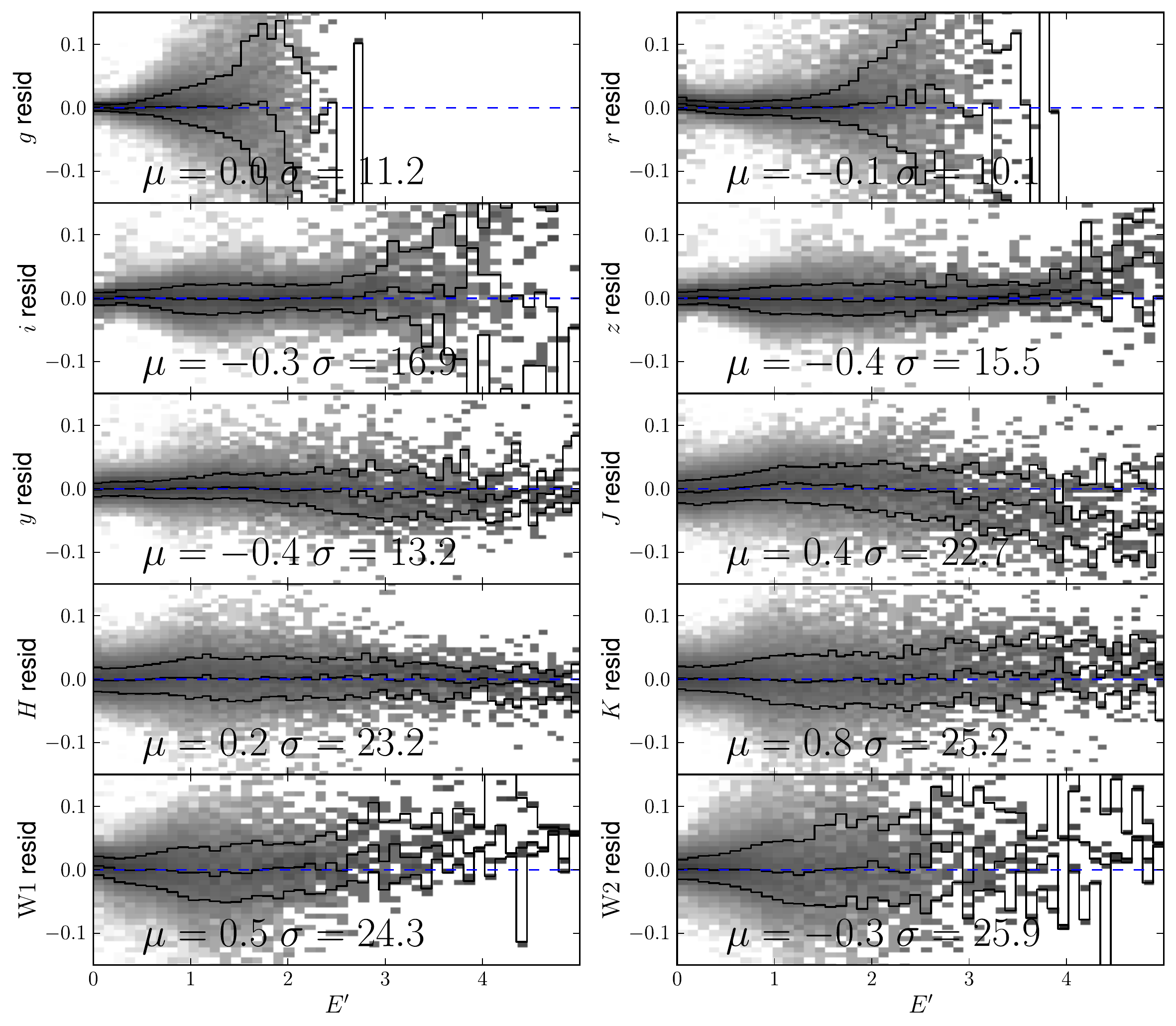}
\caption[Residuals]{
\label{fig:resid}
The residuals (data $-$ model) in the optical through infrared, as a function of the best fit reddening \Ep\ (roughly, $E(B-V)$).  The solid lines show the median residual and the 16th and 84th percentiles of the distribution as a function of \Ep.  The overall mean residual $\mu$ and rms scatter $\sigma$ are shown for each band in mmag, though we note that most of the sample has $0 < \Ep < 1$ mag.  The dispersions are essentially the same as the photometric uncertainties (10--15~mmag in PS1 $griz\yps$, 22--24~mmag in in 2MASS $JHK$ and WISE $\mathrm{W1}$ and $\mathrm{W2}$), indicating that the model is an excellent fit to the data.  At large \Ep, the residuals begin to increase significantly, due to a combination of photometric noise (in the optical bands) and variation in the extinction curve.
}
\end{figure*}

The results of our analysis are:
\begin{enumerate}
\item the mean reddening vector,
\item the way it varies,
\item the reddenings of the APOGEE targets,
\item and the intrinsic colors of APOGEE targets.
\end{enumerate}
We perform the fit of \textsection\ref{subsec:initialfit} to obtain the mean reddening vector and the intrinsic colors of APOGEE targets.  The model of Equation~\ref{eq:modelmag} proves extremely good at describing the colors of APOGEE targets in the optical through infrared.  Figure~\ref{fig:data}, Figure~\ref{fig:model}, and Figure~\ref{fig:resid} show the observed colors of APOGEE targets, their best fit model colors, and the distribution of residuals, respectively.

The model is an excellent fit to the data.  As tabulated in Table~\ref{tab:resunc}, the scatter in the residuals in a given band is typically only marginally larger than the photometric uncertainties in that band.  Stars are observed up to $\Ep \approx 5$ in bands redward of \ips, but only up to about $\Ep \approx 2.5$ in the full 10 bands, as beyond this level of extinction the stars are either saturated in \yps\ or too faint in \gps\ (for these stars, $E(g-y) > 5.5$).

\begin{deluxetable}{ccc}
\tablewidth{\columnwidth}
\tablecaption{Typical Residuals and Photometric Uncertainties
\label{tab:resunc}
}
\tablehead{
Filter & $\sigma_\mathrm{phot}/\mathrm{mmag}$ & $\sigma_\mathrm{resid}/\mathrm{mmag}$
}
\startdata
$g$ & 12 & 11  \\
$r$ & 12 & 10  \\
$i$ & 12 & 17  \\
$z$ & 12 & 16  \\
$y$ & 11 & 13  \\
$J$ & 24 & 23  \\
$H$ & 26 & 23  \\
$K$ & 22 & 25  \\
$\mathrm{W1}$ & 23 & 24 \\
$\mathrm{W2}$ & 22 & 26
\enddata
\tablecomments{
Photometric uncertanties versus root-mean-square dispersion of residuals between the data and our model.  In most bands, the scatter in the residuals is dominated by the photometric uncertainties.
}
\end{deluxetable}

\begin{figure*}[htb]
\dfplot{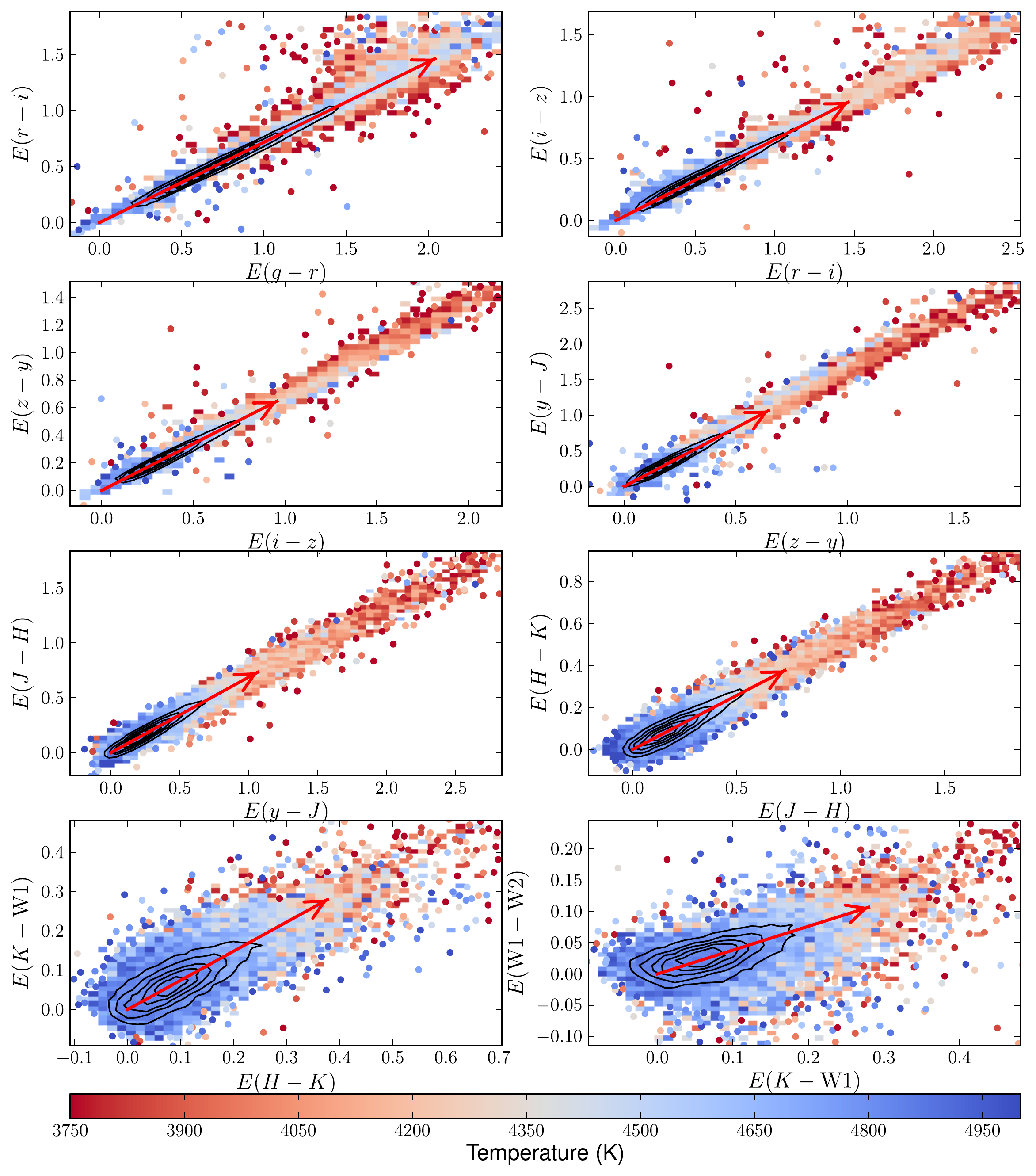}
\caption[Reddenings]{
\label{fig:reddenings}
The measured reddenings for APOGEE targets in the optical through infrared, colored by the temperature of the stars.  The Figure elements are the same as in Figure~\ref{fig:data}.  The reddenings are well described by a single fixed reddening vector, given by the red line.  The residuals from the line are not correlated with temperature, indicating that our fit for the intrinsic colors of APOGEE stars is well determined.
}
\end{figure*}

The description of reddening in terms of a single extinction curve is remarkably accurate in the optical and infrared.  We compute extinctions in each band according to
\begin{equation}
\label{eq:reddening}
\vec{r} = \vec{m} - \vec{f}(T, \mathrm{[Fe/H]}) \, ,
\end{equation}
where $\vec{r}$ gives the observed extinctions, $\vec{m}$ is the observed magnitudes, and $\vec{f}$ is the best fit function for the intrinsic colors of APOGEE targets as a function of their temperature $T$ and metallicity $\mathrm{[Fe/H]}$.  We note we have neglected $\vec{\mu}$ in Equation~\ref{eq:reddening}: since this changes only the gray component of $\vec{r}$, we are insensitive to it in our analysis.  Figure~\ref{fig:reddenings} shows the observed reddenings $\vec{r}$ in a variety of color combinations from the optical to infrared, with our mean reddening vector overplotted in red, and with points colored by their temperature.  There is little signature of a correlation between residuals and temperature in Figure~\ref{fig:reddenings}, suggesting that our fit to the intrinsic colors of APOGEE stars is accurate.

\subsection{Mean Reddening Vector}
\label{subsec:meanrf}

The first principal component of our principal component fit (\textsection\ref{subsec:pca}) is our best estimate of the mean reddening vector.  It is effectively derived from the comparison of the photometry of stars of the same temperatures and metallicities, but different reddenings---i.e., via the ``pair method,'' with the caveat that neither star is likely to be unreddened.  Table~\ref{tab:rab} shows our best fit values and their uncertainties.  These are in good agreement with the literature; see \textsection\ref{subsec:reddiscuss}.

\begin{deluxetable}{crrr}
\tablewidth{\columnwidth}
\tablecaption{The Reddening Vector and its Variation
\label{tab:rab}
}
\tablehead{
Filter & $\lambda$ & \R\ & $\dRdx$
}
\startdata

$g$ & $ 5032$ & $ 0.6537 \pm  0.0014$ & $-0.543 \pm  0.012$ \\
$r$ & $ 6281$ & $ 0.3906 \pm  0.0012$ & $ 0.034 \pm  0.020$ \\
$i$ & $ 7572$ & $ 0.2020 \pm  0.0013$ & $ 0.368 \pm  0.017$ \\
$z$ & $ 8691$ & $ 0.0787 \pm  0.0013$ & $ 0.423 \pm  0.013$ \\
$y$ & $ 9636$ & $-0.0048 \pm  0.0017$ & $ 0.382 \pm  0.025$ \\
$J$ & $12377$ & $-0.1421 \pm  0.0028$ & $ 0.141 \pm  0.034$ \\
$H$ & $16382$ & $-0.2366 \pm  0.0012$ & $-0.040 \pm  0.025$ \\
$K$ & $21510$ & $-0.2852 \pm  0.0007$ & $-0.135 \pm  0.034$ \\
$\mathrm{W1}$ & $32950$ & $-0.3213 \pm  0.0018$ & $-0.269 \pm  0.023$ \\
$\mathrm{W2}$ & $44809$ & $-0.3350 \pm  0.0015$ & $-0.363 \pm  0.036$
\enddata
\tablecomments{
The mean reddening vector \R, and the wavelengths $\lambda$ for which we expect the monochromatic extinction to be nearest these values.  The normalization and zero point of this vector is completely undetermined, and has been fixed by setting the mean of \R\ to 0 and the norm of \R\ to 1.  We also tabulate $\dRdx$, which changes the shape of the extinction curve (\textsection\ref{subsec:reddeningvariation}).  This vector is likewise mean 0 and norm 1, and we have fixed it to be perpendicular to \R.  In the language of principal component analysis, \R\ and \dRdx\ are the (normalized) loadings of the first two principal components of our analysis.
}
\end{deluxetable}

We conservatively assess the uncertainty in the reddening vector \R\ by splitting our target stars by temperature into ten equally sized subsamples and computing the root-mean-square dispersion in \R\ over these 10 subsamples.  We divide the data by temperature because temperature has a more dramatic effect on the photometry than the metallicity or gravity, especially in the optical bands where the greatest variations in intrinsic color are present.

We treat the measurements in each filter here as independent.  However, the measurements are covariant because of the free choice of mean and normalization for the vector, which is ultimately tied to the normalization and gray component degeneracies in our model (\textsection\ref{subsubsec:degeneracies}).  It is difficult to completely determine the covariance given that we expect it to be dominated by systematics, and dividing the sample into much finer bins in temperature begins to introduce significant statistical uncertainty.

We note that the mean reddening vector we derive from our principal component analysis (\textsection\ref{subsec:pca}) and from our initial fit (\textsection\ref{subsec:initialfit}) are in extremely good agreement, with typical differences of $<0.2\sigma$.

We also tabulate in Table~\ref{tab:rabslopes} our measurements of the slope of the reddening vector (i.e., $E(a-b)/E(c-d)$ for different bands $a$, $b$, $c$, and $d$) in a variety of bands.  These measurements have the advantage that they are independent of the normalization of the reddening vector and its gray component, though they also have non-trivial covariance because of shared photometric bands.  We again determine uncertainties from the dispersion over subsamples of stars of different temperatures.

\begin{deluxetable}{ccc}
\tablewidth{\columnwidth}
\tablecaption{Reddening Vector Slopes and their Variation
\label{tab:rabslopes}
}
\tablehead{
Filters & Slope & $d$Slope$/dx$
}
\startdata
$g$,$r$,$i$ & $  1.395 \pm   0.014$  &  $  -0.59 \pm    0.26$  \\ 
$r$,$i$,$z$ & $  1.531 \pm   0.013$  &  $  -2.04 \pm    0.36$  \\ 
$i$,$z$,$y$ & $  1.477 \pm   0.036$  &  $  -1.36 \pm    0.38$  \\ 
$z$,$y$,$J$ & $  0.608 \pm   0.010$  &  $  -0.77 \pm    0.29$  \\ 
$y$,$J$,$H$ & $  1.454 \pm   0.042$  &  $  -0.25 \pm    0.76$  \\ 
$J$,$H$,$K$ & $  1.943 \pm   0.020$  &  $  -0.03 \pm    0.48$  \\ 
$H$,$K$,$\mathrm{W1}$ & $  1.348 \pm   0.042$  &  $  -2.40 \pm    2.17$  \\ 
$K$,$\mathrm{W1}$,$\mathrm{W2}$ & $  2.627 \pm   0.197$  &  $  -8.18 \pm    7.70$
\enddata
\tablecomments{
The slope of the reddening vector in different bands, and the variation in that slope $d$Slope/$dx$.  The filter combination $a$,$b$,$c$ corresponds to the slope $E(a-b)/E(b-c)$.  Uncertainties are from the root-mean-square dispersion over subsets of stars with different temperatures.  This data is equivalent to that in Table~\ref{tab:rab}, except for somewhat different uncertainties due to the covariance of the measurements.
}
\end{deluxetable}

\subsection{Variation in Reddening}
\label{subsec:reddeningvariation}

\begin{figure}[htb]
\dfplot{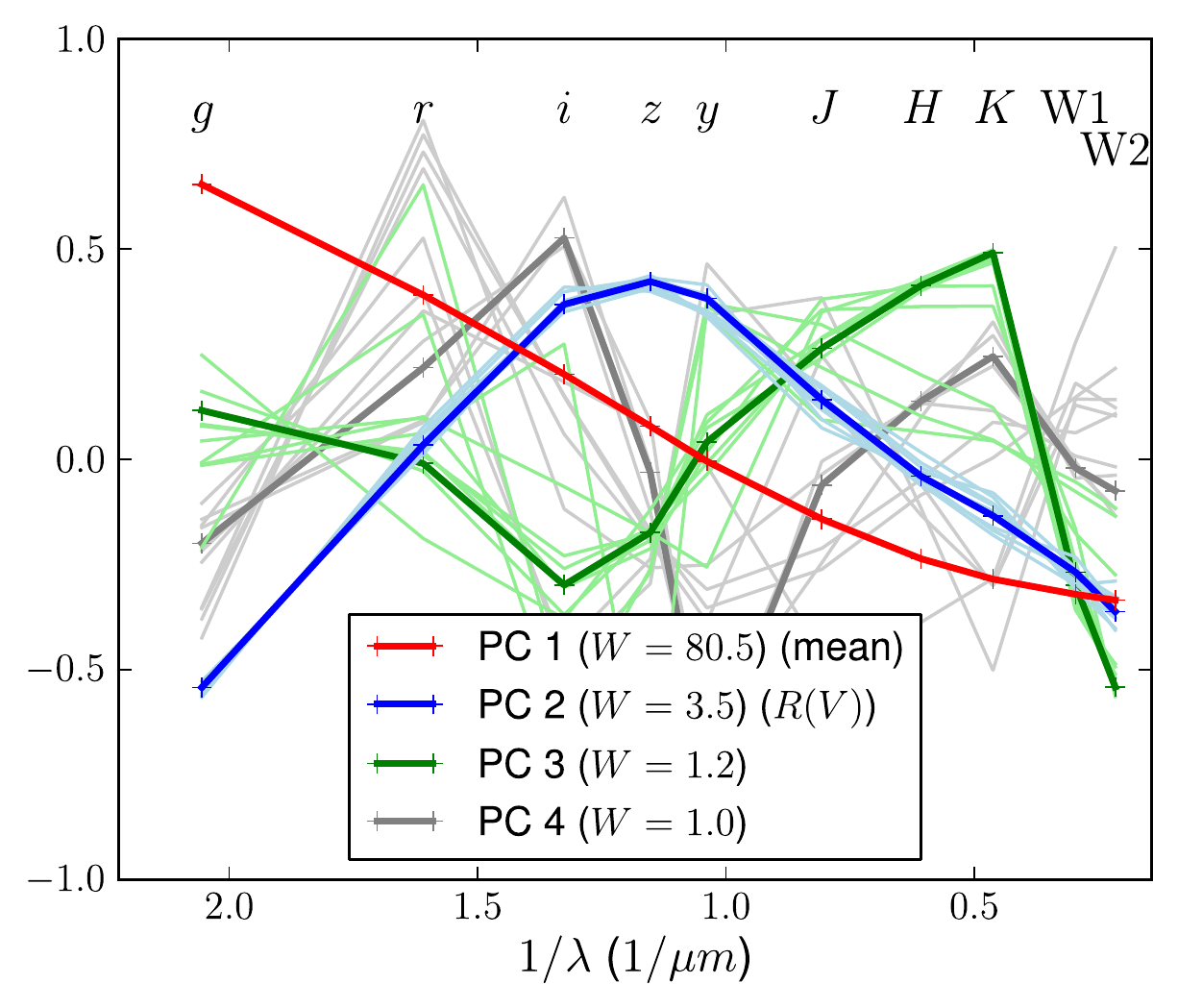}
\caption[Principal Components]{
\label{fig:pcs}
The first 4 principal components of the extinction curve (ignoring gray).  The legend gives the amount of variation in the reddenings in the direction of each principal component, relative to expectations from photometric uncertainties alone.  The top labels indicate the filter corresponding to each point.  The first principal component is essentially the mean reddening vector.  The second principal component is very similar to the effect of $R(V)$ in other formulations of the extinction curve; increasing $R(V)$ reduces the curvature of the extinction curve.  The later principal components have $W \approx 1$ in this data: they are essentially not necessary to describe the observed spectrum of a single star, though formally over the whole APOGEE sample they are statistically significant.  The thick lines show the best fit principal components from the full data set, while the thin lines show the principal components as determined from ten independent temperature subsamples.
}
\end{figure}

We assess the dimensionality of the space of reddenings by means of a version of principal component analysis that appropriately considers the uncertainties in the measurements (\textsection\ref{subsec:pca}).  We show the first four principal components in Figure~\ref{fig:pcs} (thick lines).  Thin lines in Figure~\ref{fig:pcs} show the four principal components from different temperature subsamples of the data.  Note that we force all principal components to be perpendicular to a ``gray'' reddening vector (see \textsection\ref{subsec:pca}).  Loadings for the first two principal components are tabulated in Table~\ref{tab:rab}.  

To assess the importance of these principal components, we compute the ratio of observed scatter in the component of reddening along each principal component to the expected scatter from photometric uncertainty alone.  We term this ratio $W$, and tabulate it in Table~\ref{tab:pcsig}.

\begin{deluxetable}{rrc}
\tablewidth{\columnwidth}
\tablecaption{Principal Component Significance
\label{tab:pcsig}
}
\tablehead{
\colhead{PC} \#  & \colhead{W} & \colhead{Note}
}
\startdata
 1 &  80.5 &                 mean reddening vector (see \textsection\ref{subsec:reddiscuss}) \\
 2 &   3.5 &                 $R(V)$ (see \textsection\ref{subsec:rvdiscuss}) \\
 3 &   1.2 &                 mostly consistent with noise \\
 4 &   1.0 &                 mostly consistent with noise
\enddata
\tablecomments{
The significance of the principal components: the observed scatter in the reddenings along a particular component relative to the expected scatter from photometric uncertainty alone, $W$.  We find that reddening (the first principal component) is detected, unsurprisingly, at extremely high significance in this data ($W \approx 80$).  A second principal component, similar to $R(V)$ in other extinction curve prescriptions, has $W = 3.5$.  The later principal components are detected at a significance of about $W = 1$: that is, the observed scatter in reddenings in these directions is almost completely consistent with photometric noise.
}
\end{deluxetable}

In the limit that no signal is present in the data, we expect $W=1$.  Unsurprisingly, reddening, the first principal component, is detected at high significance ($W \approx 80$).  A second principal component is detected at $W \approx 4$, which primarily acts to change the curvature of the extinction curve.  We identify this principal component with $R(V)$.  The shape of the third principal component---largely, an offset between the 2MASS and WISE photometry---leads us to attribute it small problems with the WISE photometry.  The fourth and later principal components are essentially entirely consistent with noise.  Therefore essentially all variation in the optical-infrared extinction curve detectable in our data is described by a single parameter.

We use this result to to parameterize the extinction to every star in terms of two principal components.  We define the first principal component to be \R, and the second principal component to be $\dRdx$, and tabulate these principal components in Table~\ref{tab:rab}.  We further tabulate uncertainties in $\dRdx$ via the dispersion in our measurements over different temperature subsamples of the data.  Then the extinctions $\vec{r}$ to the typical star can be expressed as 
\begin{equation}
\label{eq:redvector}
\vec{r} = (\R + x \dRdx) C \Ep \, ,
\end{equation}
where $x$ effectively determines the shape of the extinction curve toward a particular star, \Ep\ is the star's extinction according to an average curve, and $C$ is a fixed overall scale factor.  We choose $C$ so that $\Ep \approx E(B-V)$ for stars following a typical extinction curve; specifically, $C = (\R(g)-\R(r))^{-1} E(g-r)/E(B-V)$, where $E(g-r)/E(B-V) =  1.02$ is the expectation for a \citet{Fitzpatrick:1999} reddening law \citep[taken from][]{Schlafly:2011}.  Solving for $\Ep$ and $x$ for all targets with at least 7 bands of photometry and $\Ep > 0.5$, we find that the root-mean-square deviation in $x$ is $\sigma(x) = 0.023$.  In other words, roughly speaking, typical variations in the shape of the extinction curve lead to $\sim 2\%$ corrections in the reddening vector.

We note, however, that these results apply to typical stars in our sample: we do not consider the possibility that a small number of stars (say, 1\%) may have significant variations in their reddenings that are not well described by a single parameter.  In the context of the current work, these are hard to distinguish from cases where one of the surveys has provided spurious photometry.

Variation in the extinction curve is typically assessed via $R(V) = A(V)/E(B-V)$.  We cannot however directly measure $A(V)$ in this work, because we are insensitive to any gray component of the extinction curve.  However, a simple approximate proxy for $R(V)$ can be constructed from $\frac{A(g)-A(\mathrm{W2})}{A(g)-A(r)}$ (see \textsection\ref{subsec:rvlink}).  Our mean reddening vector corresponds to a proxy $R(V)$ of 3.33, and changing $x$ changes $R(V)$.  This is marginally different from the ``standard'' value of $3.1$, but given the uncertainties on the standard value and the uncertainty in translating our proxy to $R(V)$, we do not see this as problematic.

To test the robustness of these results, we have compared our heteroscedastic PCA-like analysis of \textsection~\ref{subsec:pca} with a traditional unweighted PCA analysis using the roughly 30\% of stars for which photometry is available in all ten bands.  We obtain qualitatively similar principal components for each of the first three principal components.  Quantitatively, the second, $R(V)$-like principal component is different from our preferred principal component by up to $3\sigma$ in particular bands.  We take this as remarkably good agreement given the significantly different weighting applied by our more correct analysis, where the optical photometry is weighted four times as heavily as the infrared data, due to its four times smaller variance.

\subsection{Reddenings of APOGEE stars}
\label{subsec:ebvmap}

The model provides estimates of the reddening to each of the APOGEE targets, allowing the reddening to be mapped over the footprint, as shown in Figure~\ref{fig:ebvmap}.  Observed reddenings range from $0 \leq \Ep \leq 5$~mag, though requiring a \gps\ band detection limits the range to roughly $2.5$~mag, or an \rps\ band detection to roughly $3.5$~mag.

The uncertainty of our reddening estimates to individual stars is constrained by the photometry to be less than $10$~mmag.  The dominant source of noise, however, is in the APOGEE temperatures.  Because the reddening vector and stellar locus are nearly covariant, uncertainty in temperature translates almost entirely into uncertainty in reddening (and hardly at all to uncertainty in $R(V)$ or any directions perpendicular to the reddening vector), except for stars with $T < 4000\ \mathrm{K}$.  The temperature uncertainty translates into an \Ep\ uncertainty of about 30~mmag.

\subsection{Intrinsic Optical-Infrared colors of Giants}
\label{subsec:intcolors}

\begin{figure*}[htb]
\dfplot{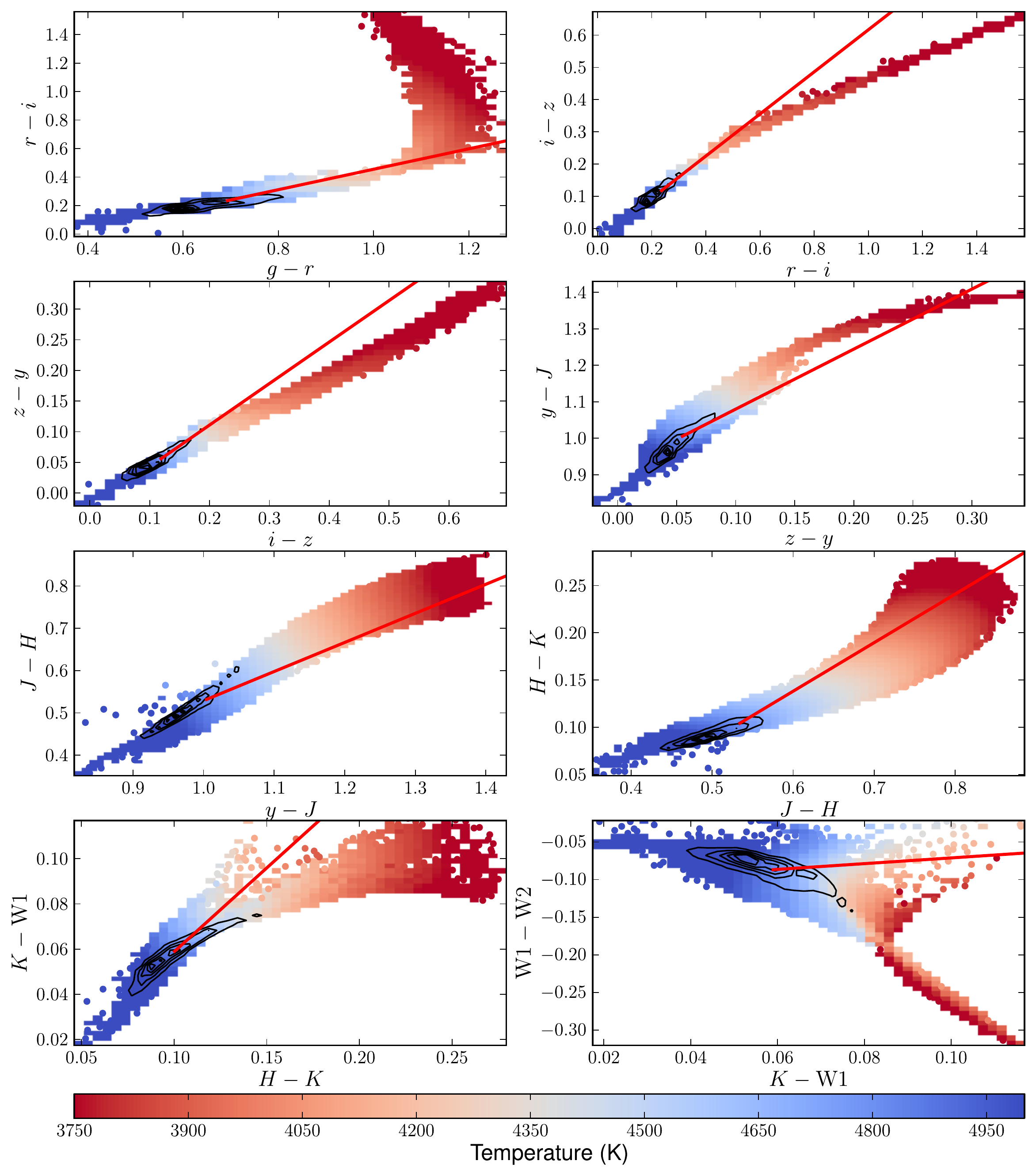}
\caption[Model Intrinsic Colors]{
\label{fig:modintrinsic}
The intrinsic colors of APOGEE stars, according to the best fit model of this work.  The Figure elements are as in Figure~\ref{fig:data}.  The colors look much as expected: the $\gps-\rps$, $\rps-\ips$ diagram (top left), for instance, shows the expected sharp bend at $\gps-\rps = 1.2$~mag typical of PS1 filters.  The colors in all filter combinations are determined primarily by temperature, though the distribution of points is also significantly broadened by colors' dependence on metallicity; iso-color lines show the variation in color due to varying metallicity at fixed temperature.  The $\rps-\ips$, $\ips-\zps$ color-color diagram is found to be especially close to being a single parameter family; meanwhile in the infrared, metallicity dramatically broadens the stellar locus.
}
\end{figure*}

Our analysis also gives the intrinsic optical-infrared colors of giant and subgiant stars as a function of their APOGEE parameters.  Figure~\ref{fig:modintrinsic} shows color-color diagrams of the intrinsic colors our model predicts for each of the APOGEE stars.  The colors look as expected: in $gr\ips$, the stellar locus features a prominent bend at $\gps-\rps \approx 1.2$; the other optical bands are essentially linear and relatively one-dimensional (little dependence of color on metallicity).  In the infrared, metallicity begins to play a larger role relative to temperature than in the optical.  The near collinearity of the reddening vector and the stellar locus in the optical largely disappears in the WISE bands, as recognized by \citet{Majewski:2011}, which motivated the APOGEE dereddening and target selection algorithm of \citet{Zasowski:2013}.

As discussed in \textsection\ref{subsubsec:degeneracies}, our intrinsic color measurements are subject to a perfect degeneracy.  Since we do not assume we know where zero reddening lies, the entire set of colors can be shifted along the reddening vector without having any effect on the goodness of fit.  Worse, the analysis essentially works by comparing the reddenings of stars with similar temperatures and metallicities, and so in fact stars of different temperatures and metallicities can be shifted by different amounts along the reddening vector without affecting the analysis.

\begin{figure}[htb]
\dfplot{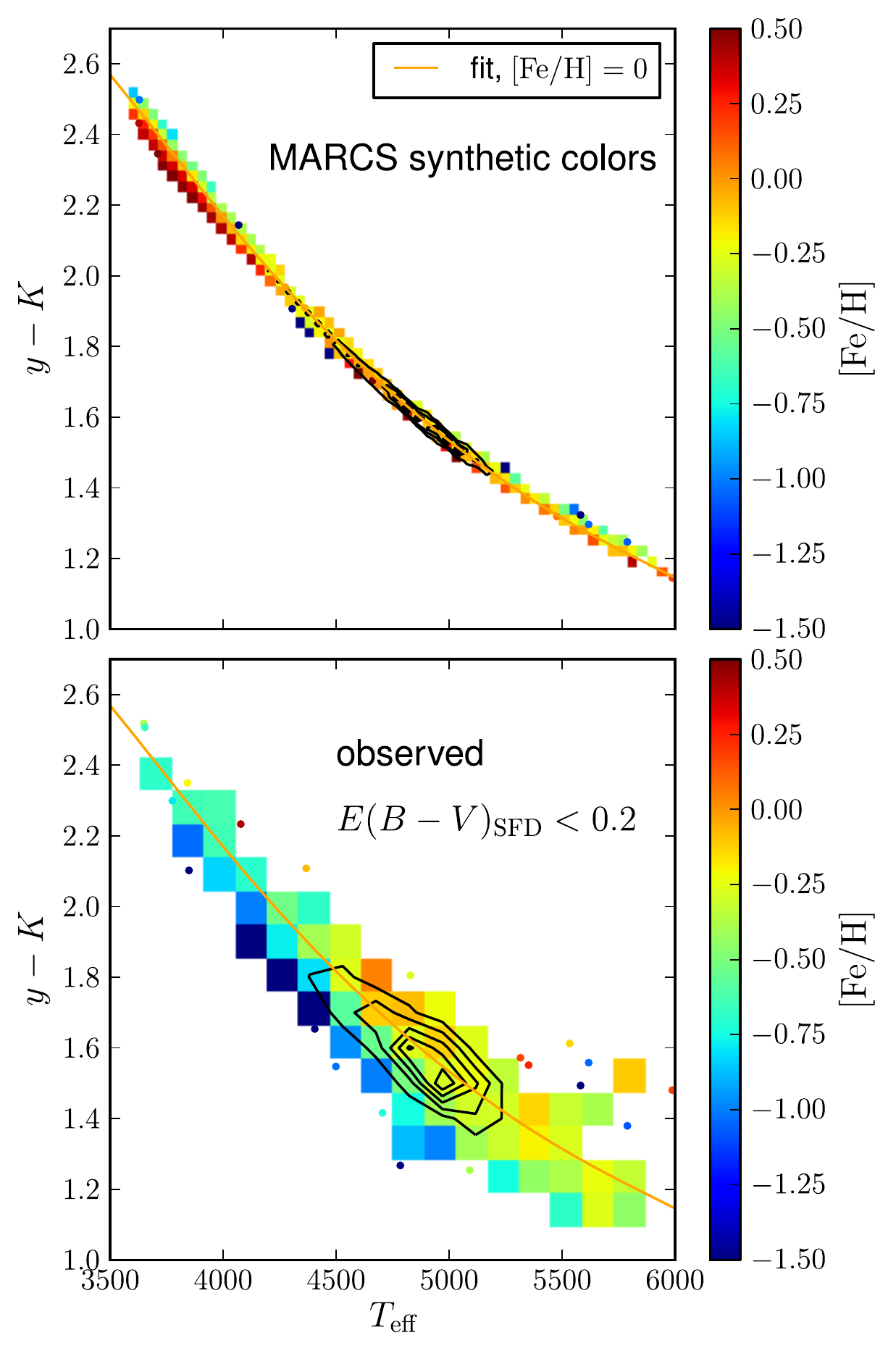}
\caption[Intrinsic $y-k$ color]{
\label{fig:intyk}
The intrinsic $\yps-K$ color of APOGEE stars, as synthesized from MARCS model spectra (top) and as observed for $E(B-V)_\mathrm{SFD} < 0.2$~mag stars after dereddening according to SFD.  The results of our fit to the MARCS predictions, for $\mathrm{[Fe/H]} = 0$, is shown as the solid line in each panel.  There is good overall agreement between the MARCS predictions and the observations, though the MARCS colors predict less variation in color with metallicity than is observed.  Points are colored by metallicity, and a fit to the MARCS colors at $\mathrm{[Fe/H] = 0}$ is is shown in each panel as a solid line.  The contours show the density of stars.
}
\end{figure}

To address this shortcoming, we fix the intrinsic $\yps-K$ color to be a function of our choice, intended to be the true $\yps-K$ color of unreddened stars.  We choose to force our intrinsic $\yps-K$ colors to agree with the predictions of the MARCS model grid of \citet{Gustafsson:2008}, which were found to be good predictors of broadband colors by \citet{Edvardsson:2008} and \citet{Casagrande:2014}.  It is challenging to fully assess the accuracy of these predictions with the APOGEE stars themselves.  The most straightforward approach is to compare the $\yps-K$ color of APOGEE targets in regions of low reddening with the predictions from model spectra.  The results of such a comparison are shown in Figure~\ref{fig:intyk}.

The top panel of Figure~\ref{fig:intyk} shows the predicted $\yps-K$ color of APOGEE stars as a function of their temperature, with points colored by their metallicity.  In the models, the $\yps-K$ color has very little dependence on metallicity.  The solid line shows our fit to these synthetic colors, at $\mathrm{[Fe/H]} = 0$.  The fit is extremely good; the root-mean-square difference between our fit and the actual prediction is 1.3~mmag.  The observed colors, dereddened according to \citet[SFD]{Schlegel:1998}, are shown in the bottom panel, using only stars for which $E(B-V)_\mathrm{SFD} < 0.2$.  We have again shown the $\mathrm{[Fe/H]} = 0$ fit from the top panel, to illustrate the good agreement in general between the synthetic and observed magnitudes.  Still, the fit is not perfect: there is a noticeable trend in $\yps-K$ color with metallicity in the observed colors, which is absent or much reduced in the synthetic colors.  Moreover, below 4000~K there are very few unreddened APOGEE targets, and none of these are solar metallicity or above, making it impossible to test the accuracy of the fit in this region.  Still, the fit is at least consistent with the limited data available there.

We can dramatically increase the number of stars available for this test by adopting a more permissive cut on $E(B-V)_\mathrm{SFD}$, essentially using stars closer to the plane that are more likely to be distant solar metallicity giants.  However, SFD is known to be problematic at low latitudes, especially in the inner Galaxy \citep[e.g.,][]{Schlafly:2014b}.  Extending Figure~\ref{fig:intyk} to $E(B-V)_\mathrm{SFD} < 0.5$~mag renders the observed metallicity trend invisible, presumably due to systematic overprediction of the reddening in the plane in SFD, possibly due to the fact that the APOGEE targets may not be behind the entire dust column.   

Due to the degeneracies, our choice of $\yps-K$ color does not affect the reddening vector we derive, or its variation.  Accordingly, we simply fix $\yps-K$ to the value we fit from the MARCS stellar models.  Errors in this choice lead only to small variations in the \Ep\ we infer to individual stars.

\subsection{Quality of the Fit}
\label{subsec:fitquality}

\begin{figure*}[htb]
\dfplot{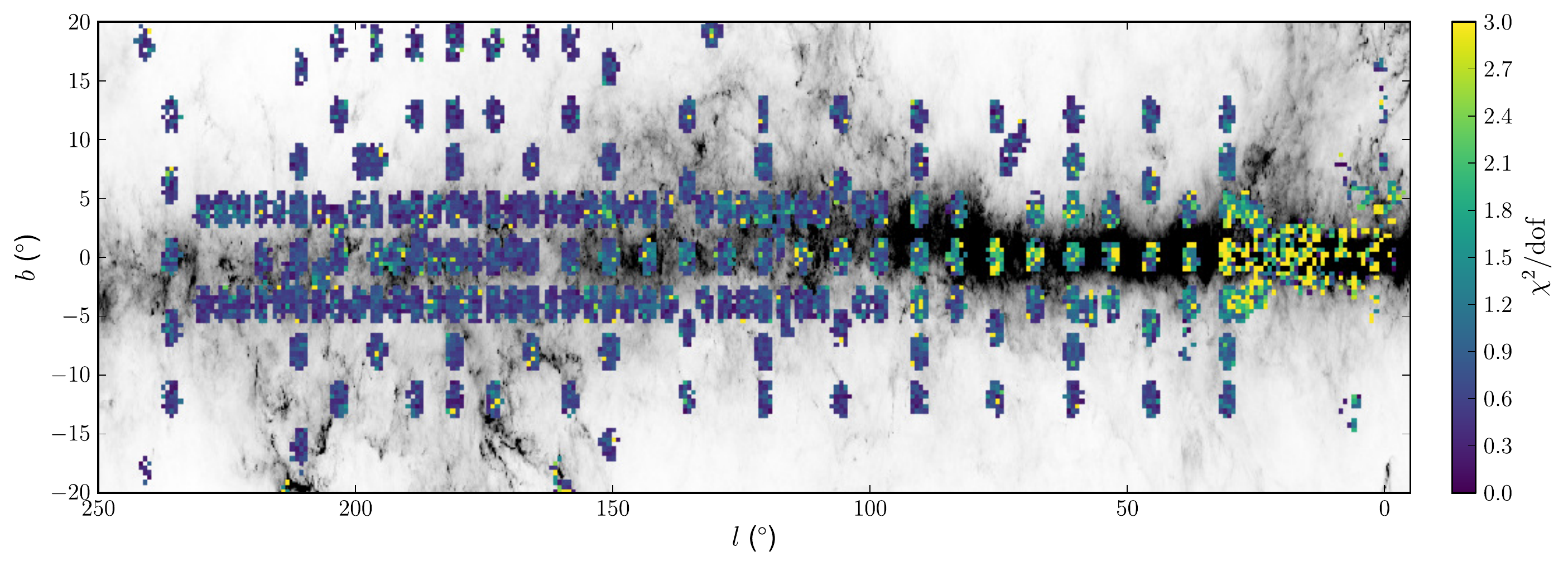}
\caption[$\chi^2/\mathrm{dof}$ of APOGEE Targets]{
\label{fig:chi2dof}
$\chi^2$ per degree of freedom for APOGEE targets.  Throughout most of the disk, we obtain a $\chi^2/\mathrm{dof}$ of 0.77, suggesting our uncertainties are very slightly overestimated.  In the inner disk and especially toward the Galactic center, $\chi^2/\mathrm{dof}$ becomes significantly larger and more variable.  Much of the contribution in the Galactic center comes from the WISE W1 and W2 bands, suggesting that crowding and blending may be problematic.  This may also be a signal that the dust extinction curve is more variable in these directions, though we do not find conclusive signatures of that possibility.
}
\end{figure*}

We find in \textsection\ref{subsec:reddeningvariation} that the extinction curve to our stars can be well parameterized in terms of two principal components.  We can assess the quality of the fit by modeling the observed extinctions $\m - \vec{f}(T, \mathrm{[Fe/H]})$ as a sum of a gray component, the mean reddening vector \R, and \dRdx\ (i.e., an $R(V)$-like component).  We then compute the $\chi^2$ per degree of freedom for each star.  We show the spatial distribution of $\chi^2$ per degree of freedom in Figure~\ref{fig:chi2dof}.

The mean $\chi^2$ per degree of freedom in Figure~\ref{fig:chi2dof} is 0.77, slightly smaller than the expected value of one, suggesting our uncertainties are slightly overestimated.  As Figure~\ref{fig:chi2dof} makes clear, however, the inner Galaxy and especially the Galactic bulge feature significantly larger $\chi^2$ per degree of freedom than the entire outer Galaxy, which is relatively featureless.

There are a few reasons for the large $\chi^2$ per degree of freedom in the inner Galaxy.  The first is that we have included in the computation of $\chi^2$ only the contribution from the photometric uncertainties, and neglected the contribution of the spectroscopic uncertainties in $T$ and $\mathrm{[Fe/H]}$.  For most stars, the spectroscopic uncertainties affect $\chi^2$ negligibly, because the reddening vector and stellar locus are well aligned, and any error in $T$ can therefore be absorbed into the extinction in the model.  For the coldest stars ($ T \lesssim 4000~\mathrm{K}$), however, the reddening vector becomes nearly perpendicular to the stellar locus, and colors become more sensitive to temperature.  This can lead to uncertainties in $T$ dominating the total uncertainties for these stars.  In principle we could build the temperature uncertainty into the analysis, but we have found that doing so makes no difference for the reddening vector we derive, and so we have neglected this contribution.  Because essentially all of the $T < 4000~\mathrm{K}$ stars reside in the inner Galaxy, this leads to elevated $\chi^2$ there.

However, even when excluding all cold stars we still find elevated $\chi^2$ per degree of freedom in the inner Galaxy.  The primary driver seems to be residuals in the WISE bands, though additionally excluding these bands does not fully resolve the problem.

We have not been able to fully understand the cause of the large $\chi^2$ per degree of freedom in the inner Galaxy.  One explanation could be that the extinction curve is more variable or differently variable in the inner Galaxy \citep[e.g.,][]{Nataf:2013, Nataf:2015}.  However, applying our analysis only to stars with $|l| < 60\degree$ and $|b| < 5\degree$ produces similar principal components as when using our full data set.  In particular, the third principal component continues to bear signs of a WISE-2MASS offset, which we find unlikely to have a physical origin.

We conclude that over the great majority of the sky, our two component extinction curve model provides an excellent description of the data.  In the inner Galaxy, this model clearly fails to account for the data completely.  Nevertheless, even in this region, the two component model describes the majority of the variation in the extinction curve.

\section{Discussion}
\label{sec:discussion}

In this section, we discuss the implications of our results and compare with similar measurements from the literature.  We
\begin{enumerate}
\item discuss our one-parameter family of extinction curves,
\item compare our extinction curve with the literature,
\item link our observed variation with $R(V)$,
\item study the variation of the extinction curve through the Galaxy, 
\item study correlations between the dust emission and extinction curves,
\item discuss measurements of the ``gray'' component of the extinction, and
\item compare our intrinsic colors with models.
\end{enumerate}

\subsection{The Extinction Curve}
\label{subsec:rvdiscuss}

We find in \textsection\ref{subsec:reddeningvariation} that the reddening to a star in the optical and infrared can be parameterized by two numbers: essentially, the amount of reddening to the object, and a second parameter that slightly alters the direction of the reddening vector.  This is the same conclusion that CCM made, and popularized the idea that the extinction curve can be parameterized by $R(V) = A(V) / E(B-V)$.  

This work distinguishes itself from CCM in that that work focused largely on the UV, while we are concerned entirely with the optical-infrared extinction curve.  Moreover, the UV extinction curve is in detail not a single parameter family---the work of \citet{Fitzpatrick:1990} uses six parameters to describe the UV extinction curve, and the CCM curve was found to describe the majority of the variability among those parameters.  However, we find that the bulk of the APOGEE data provide no support for using more than a single parameter family to describe the extinction in the optical through infrared, with the possible exception of the inner Galaxy, where we can draw no firm conclusions (\textsection\ref{subsec:fitquality}).

\begin{figure*}[h!tb]
\dfplot{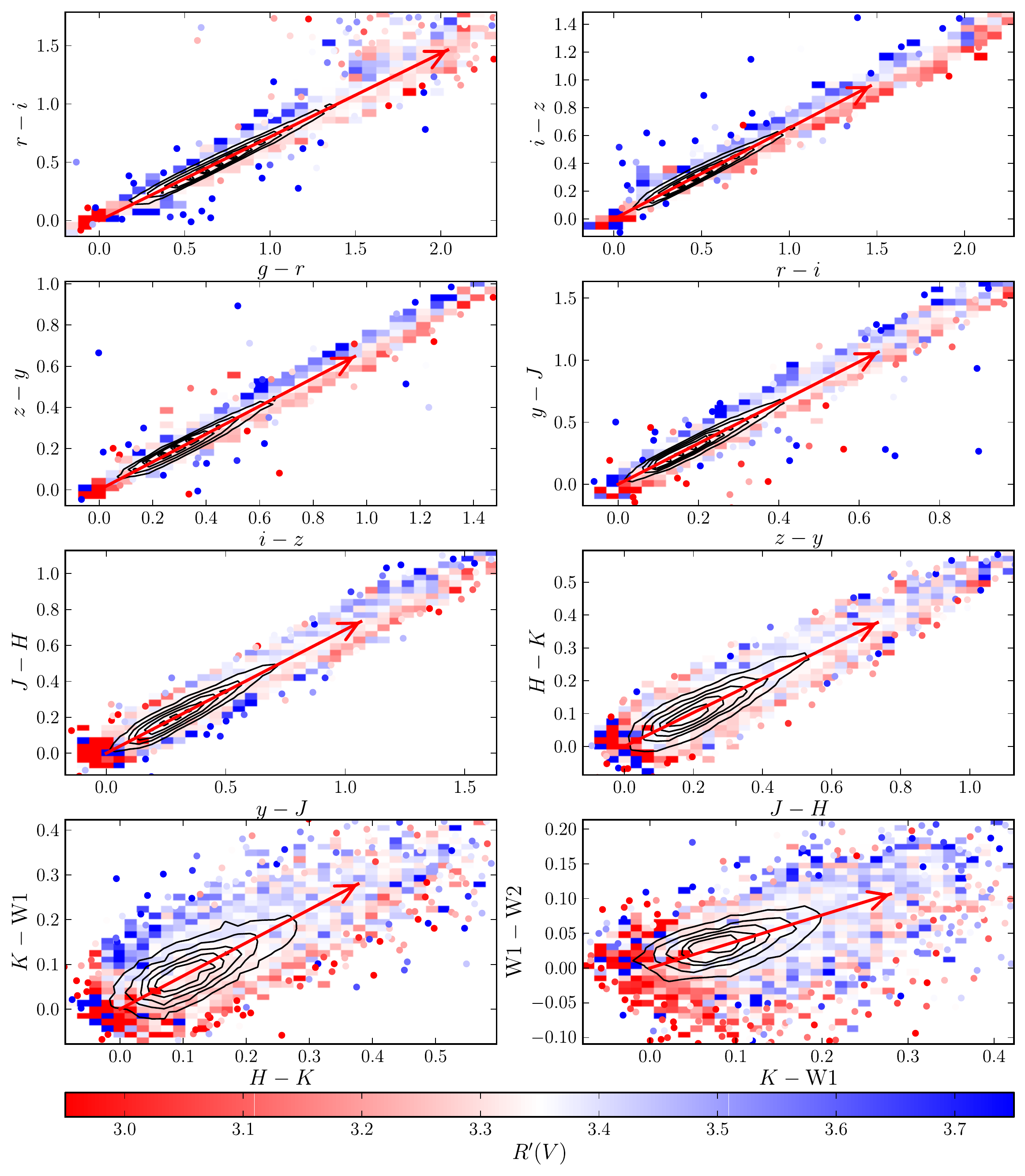}
\caption[Reddenings versus $R(V)$]{
\label{fig:reddeningsrv2}
Reddenings of APOGEE stars in different color combinations, colored by our principal-component inferred \RpV.  The Figure elements are as in Figure~\ref{fig:data}.  The clearest changes in slope with \RpV\ are in the bluer bands (\gps\ through \zps), while in the infrared the slope variation is relatively small.
}
\end{figure*}

We find that the curvature of the extinction curve increases with decreasing $R(V)$ throughout the optical and infrared.  This can be directly seen in Figure~\ref{fig:reddeningsrv2}, where we show observed reddenings of stars, colored by \RpV\ as estimated from our principal component analysis, for stars with photometry in at least 9 bands.  In the optical bands, there is a clear association between the slope of the reddening vector and \RpV.  Meanwhile redward of $J$, the trend becomes harder to detect.  However, we note that the work of \citet{Zasowski:2009}, which had access to stars of significantly larger reddening, found that the longest wavelengths had the most significant variations.

\subsection{The Extinction Curve Compared with Past Measurements}
\label{subsec:reddiscuss}
In this subsection, we compare our mean extinction curve with past measurements of the optical-infrared extinction curve from broad-band photometry, and with standard parameterizations of the extinction curve.  Our measurements rely on a large sample of highly reddened stars with precise, homogeneous stellar parameters and photometry, covering much of the Galactic plane.  We therefore believe our measurements of the extinction curve supersede earlier works.  

\subsubsection{Comparison with Photometric Measurements}
\label{subsubsec:redphotometric}

The slope of the reddening vector has been measured by a large number of authors in a variety of photometric bands in the optical through infrared.  We show in Table~\ref{tab:litcomp} our $E(a-b)/E(c-d)$ in different combinations of bands, as compared with a selection of measurements from the literature.

The bands we study are not always a perfect match for the bands considered in other works.  To translate measurements from one set of bands into another, we use an F99 extinction curve to compute predictions for the expected ratios $A(a)/A(b)$.  Because the slope of the reddening vector also depends slightly on the intrinsic spectrum of a star and its reddening, we further additionally slightly correct the measurements of \citet{Schlafly:2011} and \citet{Yuan:2013} to account for this difference.  To make this adjustment, we assume in in Table~\ref{tab:litcomp}, we assume that these two sets of measurements studied lightly reddened ($E(B-V) \approx 0.1)$) blue main sequence stars (7000~K), while we study reddened ($E(B-V) \approx 0.65$) red giants (4500~K).  However, the choice of spectrum usually makes a difference of $<0.01$ in $E(a-b)/E(c-d)$.

\begin{deluxetable*}{cllclc}
\tablewidth{\textwidth}
\tablecaption{Mean Reddening Vector Compared to Past Measurements
\label{tab:litcomp}
}
\tablehead{
\colhead{Filters} & \colhead{This work} & \colhead{Lit. conv.} & \colhead{Orig. filters} & \colhead{Orig. value} & \colhead{Ref.}
}
\startdata
$E(gP-rP)/E(rP-iP)$ & $1.395 \pm 0.013$ & $1.449 \pm 0.049$ & $E(gS-rS)/E(rS-iS)$ & $1.695 \pm 0.057$ & Schlafly+2011 \\
$E(rP-iP)/E(iP-zP)$ & $1.531 \pm 0.012$ & $1.578 \pm 0.041$ & $E(rS-iS)/E(iS-zS)$ & $1.299 \pm 0.034$ & Schlafly+2011 \\
$E(gP-rP)/E(rP-iP)$ & $1.395 \pm 0.013$ & $1.410 \pm 0.038$ & $E(gS-rS)/E(rS-iS)$ & $1.650 \pm 0.044$ & Yuan+2013 \\
$E(rP-iP)/E(iP-zP)$ & $1.531 \pm 0.012$ & $1.695 \pm 0.041$ & $E(rS-iS)/E(iS-zS)$ & $1.395 \pm 0.034$ & Yuan+2013 \\
$E(iP-zP)/E(zP-J2)$ & $0.558 \pm 0.008$ & $0.558 \pm 0.015$ & $E(iS-zS)/E(zS-J2)$ & $0.768 \pm 0.020$ & Yuan+2013 \\
$E(zP-J2)/E(J2-H2)$ & $2.338 \pm 0.076$ & $2.434 \pm 0.105$ & $E(zS-J2)/E(J2-H2)$ & $2.154 \pm 0.093$ & Yuan+2013 \\
$E(J2-H2)/E(H2-K2)$ & $1.943 \pm 0.019$ & $1.627 \pm 0.063$ &                     & $1.625 \pm 0.063$ & Yuan+2013 \\
$E(H2-K2)/E(K2-W1)$ & $1.348 \pm 0.040$ & $1.318 \pm 0.093$ &                     & $1.333 \pm 0.094$ & Yuan+2013 \\
$E(K2-W1)/E(W1-W2)$ & $2.627 \pm 0.187$ & $4.653 \pm 1.440$ &                     & $4.615 \pm 1.428$ & Yuan+2013 \\
$E(gP-iP)/E(J2-K2)$ & $3.157 \pm 0.066$ & $3.676 \pm 0.214$ & $E(VL-IL)/E(J2-K2)$ & $2.913 \pm 0.170$ & Nataf+2013 \\
$E(gP-rP)/E(rP-iP)$ & $1.395 \pm 0.013$ & $1.443          $ & $E(gS-rS)/E(rS-iS)$ & $1.625          $ & Davenport+2014 \\
$E(rP-iP)/E(iP-zP)$ & $1.531 \pm 0.012$ & $1.283          $ & $E(rS-iS)/E(iS-zS)$ & $1.043          $ & Davenport+2014 \\
$E(iP-zP)/E(zP-J2)$ & $0.558 \pm 0.008$ & $0.733          $ & $E(iS-zS)/E(zS-J2)$ & $1.000          $ & Davenport+2014 \\
$E(zP-J2)/E(J2-H2)$ & $2.338 \pm 0.076$ & $2.852          $ & $E(zS-J2)/E(J2-H2)$ & $2.556          $ & Davenport+2014 \\
$E(J2-H2)/E(H2-K2)$ & $1.943 \pm 0.019$ & $1.500          $ &                     & $1.500          $ & Davenport+2014 \\
$E(H2-K2)/E(K2-W1)$ & $1.348 \pm 0.040$ & $1.000          $ &                     & $1.000          $ & Davenport+2014 \\
$E(K2-W1)/E(W1-W2)$ & $2.627 \pm 0.187$ & $1.500          $ &                     & $1.500          $ & Davenport+2014 \\
$E(J2-H2)/E(H2-K2)$ & $1.943 \pm 0.019$ & $2.000 \pm 0.050$ &                     & $2.000 \pm 0.050$ & Zasowski+2009 \\
$E(J2-H2)/E(H2-K2)$ & $1.943 \pm 0.019$ & $1.780 \pm 0.008$ &                     & $1.780 \pm 0.008$ & Wang+2014 \\
$E(J2-H2)/E(H2-K2)$ & $1.943 \pm 0.019$ & $1.778 \pm 0.154$ &                     & $1.778 \pm 0.154$ & Indebetouw+2005
\enddata
\tablecomments{
The extinction curve of this work compared with the literature.  The first column gives the reddening vector slope of interest, and the second column gives our measurement of it.  The third column gives measurements and uncertainties from the literature, converted to be in the same filters as our measurements, when applicable.  The fourth column gives the original filters the literature measurement was made in, if different from ours, and the fifth column gives the original values of the measurement from the literature.  Finally, the sixth column gives the literature reference.  Bandpasses from the \PS, SDSS, 2MASS, and Landolt systems are denoted with the characters P, S, 2, and L, respectively.  The measurements of \citet{Schlafly:2011} and \citet{Yuan:2013} have been further adjusted to account for the fact that these works studied intrinsically hotter, less reddened stars than the stars targeted in APOGEE; see text.
}
\end{deluxetable*}

The agreement between our measurements and the literature is good.  We agree with \citet{Schlafly:2011} in $E(\gps-\rps)/E(\rps-\ips)$ and $E(\rps-\ips)/E(\ips-\zps)$ to within about $1\sigma$.  Our agreement with \citet{Yuan:2013} is acceptable, though there is a more than $4\sigma$ disagreement in $E(r-i)/E(i-z)$ and in $E(J-H)/E(H-K)$.  We also measure $E(K-\mathrm{W1})/E(\mathrm{W1}-\mathrm{W2})$ to be nearly twice as large as \citet{Yuan:2013}, though their measurement was uncertain due to the low reddenings available in their work.  We are about $2\sigma$ discrepant from the work of \citet{Nataf:2013}, though we note that the spread in $E(V-I)/E(J-K)$ reported there is attributed to real variation in the extinction curve rather than uncertainty, making a $2\sigma$ offset very significant.  We find in general poor agreement with the results of \citet{Davenport:2014}, possibly due to a systematic effect arising from adopting an intrinsic color relation best suited to dwarfs to find the reddenings of likely giants.

The work of \citet[W14]{Wang:2014} makes a measurement of $E(J-H)/E(H-K)$ from similar data to ours: APOGEE spectroscopy combined with 2MASS data.  Our values, however, are in conflict: we find $E(J-H)/E(H-K) = 1.943$ as compared with their $1.780$, a huge discrepancy given our estimated uncertainty of $0.019$ and their quoted uncertainty of $0.008$.  This difference is especially disturbing given the largely identical data adopted by our work and that of W14.  To determine the source of the discrepancy, we repeated our analysis, this time including only 2MASS data and excluding PS1 and WISE data.  We could reproduce their results only if we also adopted their equations for the intrinsic colors of giant stars as a function of their temperature (Equations 1, 2, and 3 in W14).  This highlights the importance of the intrinsic colors to this analysis: because the reddest and most metal rich APOGEE stars tend to lie at the greatest distances toward the Galactic center, and are accordingly the most reddened, a systematic trend in intrinsic color with temperature can masquerade as a different reddening vector.  In our analysis, we allow the intrinsic colors to be fit simultaneously with the reddening vector, and obtain a fit with significantly better $\chi^2$ than we obtain with the W14 intrinsic color relation.  We conclude that our value is much more likely to be correct, and note that it is in good agreement with the work of \citet{Zasowski:2009}.

\subsubsection{Comparison with Existing Extinction Curve Parameterizations}

We can additionally compare our results with parameterized extinction curves from the literature: we consider the curves of \citet[CCM]{Cardelli:1989}, \citet[F99]{Fitzpatrick:1999}, \citet[FM04]{Fitzpatrick:2004}, \citet[FM09]{Fitzpatrick:2009}, and \citet[M14]{MaizApellaniz:2014}.  The CCM and M14 extinction curves are not defined redward of 33333~\AA; we extend these curves redward using the FM09 curve.  Similarly, the FM09 extinction curve was developed using data redward of 6000~\AA; we extend it blueward of 6000~\AA\ using F99.

We note that each of these extinction curves is a one-parameter family.  In the case of all but the FM09 extinction curve, the controlling parameter is referred to as $R(V)$; in the case of FM09, it is referred to as $\alpha$.  FM09 has a second parameter, $R(V)$, that does not change the shape of the extinction curve redward of $V$; we ignore it here.

We compute predictions of the slope of the reddening vector we should observe from extinction curves by integrating the appropriate filter bandpasses over the MARCS synthetic spectrum of a 4500~K star with $\log g = 2.5$ and solar metallicity, typical of the APOGEE sample.  Since the APOGEE sample has few unextinguished stars, the slope we observe is actually the slope of the reddening vector at $E(B-V) \approx 0.65$ rather than at $E(B-V) \approx 0$.  To account for this, we compute $dm_b/dA$ for small variations of the extinction $dA$ about $E(B-V) = 0.65$, where $m_b$ is the observed magnitude in the bandpass $b$.

This procedure generates $dm_b/dA$ in each bandpass $b$, but in APOGEE we are insensitive to the normalization of the extinction and to any gray component of the extinction.  So to compare with Table~\ref{tab:rab}, we fit extinction curves to the measurements as
\begin{equation}
\R_{b, \mathrm{obs}} = C \cdot d\m_b/dA + D \, ,
\end{equation}
in a least squares sense, using the uncertainties given in Table~\ref{tab:rab} and neglecting any covariance in the uncertainties.

Performing the fit to each family of extinction curves, the FM09 extinction curve is by far most consistent with our measurements ($\chi^2 = 25.5$).  The F99 extinction curve is next best ($\chi^2 = 92.7$), followed by the FM04 extinction curve ($\chi^2 = 202.5$).  The CCM and M14 extinction curves are strongly disfavored ($\chi^2 = 633.1$ and $\chi^2 = 861.0$).  These $\chi^2$ are all on 7 degrees of freedom, except for FM09, which is on 6 degrees of freedom, since we have excluded the $\gps$ band, which is outside the region where the FM09 prescription was developed.

\begin{figure}[htb]
\dfplot{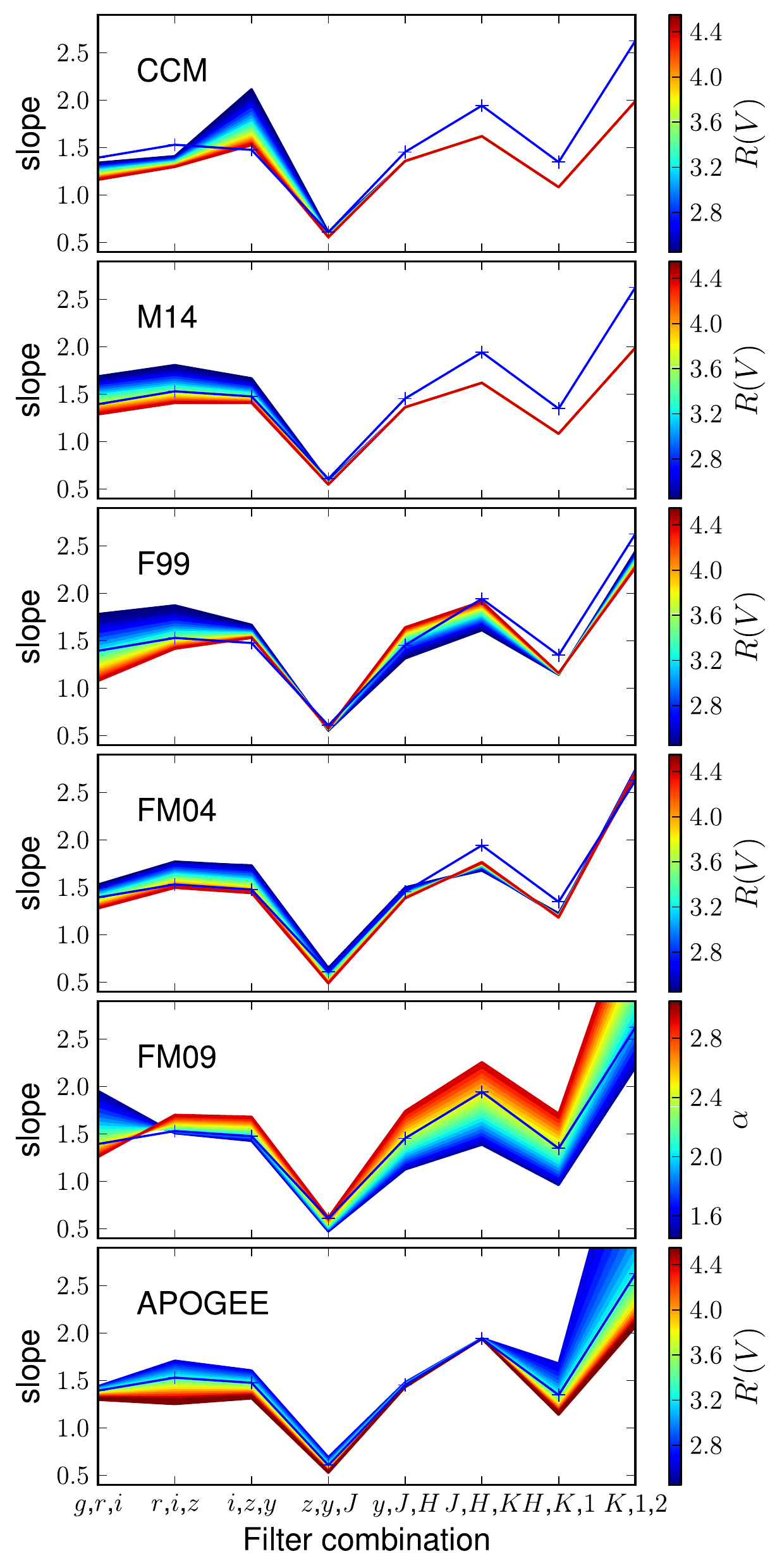}
\caption[Extinction Curves compared with APOGEE Measurements]{
\label{fig:redrvcurves}
Different parameterizations of the extinction curve compared with our measurements from APOGEE.  We consider the CCM, M14, F99, FM04, and FM09 extinction curves.  The $y$-axis shows the slope of the reddening vector in a particular combination of filters, as indicated on the $x$-axis.  The filter combination $g$,$r$,$i$ corresponds to the slope $E(g-r)/E(r-i)$, and analogously for the other filter combinations.  Different colors correspond to different values of $R(V)$ or $\alpha$.  The blue line with crosses shows the APOGEE measurements.  The bottom panel shows our APOGEE-determined extinction curve.  The FM09 extinction curve best matches our mean extinction curve, though its prediction for the variation in the extinction curve is significantly different from what we measure.
}
\end{figure}

This simple description of which extinction curve works best hides many important details of the extinction curves.  The M14 extinction curve, for instance, provides an extremely good fit at $R(V) = 3.7$ in the PS1 bands ($\chi^2 = 1.0$ on 2 degrees of freedom).  However, the IR extension of M14 follows CCM, which poorly matches our measurements, leading to high $\chi^2$ overall.  We compare the behavior of the various extinction curves with that which we find from APOGEE in Figure~\ref{fig:redrvcurves}, which shows the slope of the reddening vector in different photometric bands for a variety of extinction curves families.

The bottom panel of Figure~\ref{fig:redrvcurves} shows that we observe variation in the slope of the reddening vector across the optical and infrared, with significantly reduced variation in the near-infrared relative to the optical.  Associating this variation with $R(V)$ according to Equation~\ref{eq:rvproxy}, we find that the slope of the reddening vector increases with decreasing $R(V)$ throughout the optical and infrared.  The general trend of the variation is qualitatively similar to that of all curves except the FM09 and F99 curves: the optical bands show more variability than the infrared bands.  However, in detail none of the extinction curves come especially close to our observations.

Restricting to bands blueward of $J$, the M14 family provides both a good description of the mean extinction curve and its variation with $R(V)$.  It however qualitatively disagrees with our measurement for the amount of variation in $E(g-r)/E(r-i)$ relative to $E(r-i)/E(i-z)$: we find more significant variation in the latter than the former.

\begin{figure}[htb]
\dfplot{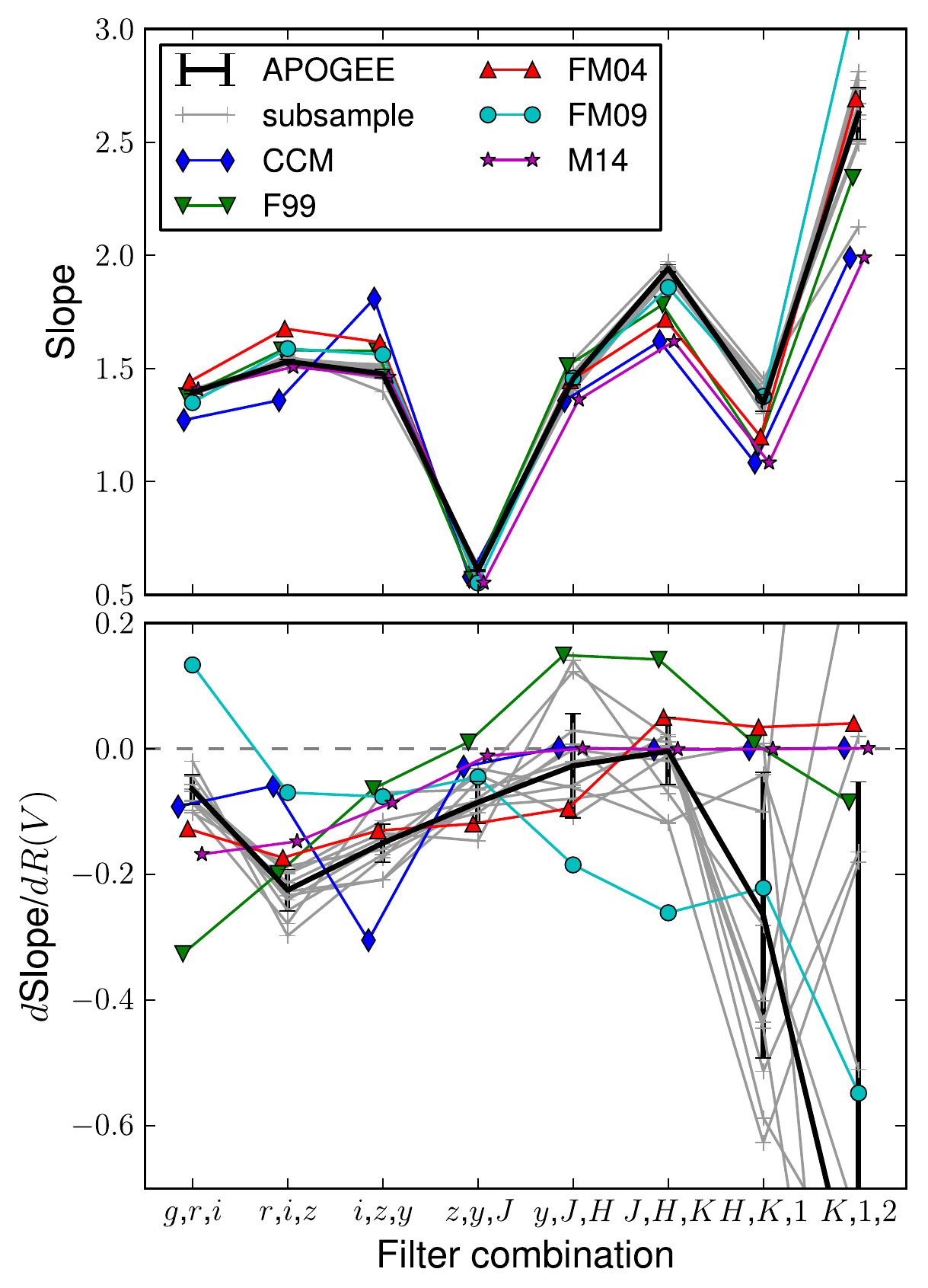}
\caption[Extinction curves compared with APOGEE Summary]{
\label{fig:redrvcurvessummary}
Comparison of various extinction curves and our measurements from APOGEE.  The top panel shows predictions for the slope of the extinction curve in different filter combinations (see Figure~\ref{fig:redrvcurves} caption for details), and the bottom panel shows the derivative of the slope with respect to $R(V)$ (or $\alpha$ in the case of FM09, with an arbitrary rescaling).  The black line with error bars shows our APOGEE measurements, while the thin gray lines show our measurements for subsamples of the data with different temperatures.  The FM09 extinction curve provides the best match, but quantitatively no extinction curve agrees especially well with our measurements of the variation of the shape of the extinction curve.
}
\end{figure}

We summarize these results in Figure~\ref{fig:redrvcurvessummary}.  The upper panel compares the predictions for the slope of the reddening vector in different filter combinations for various extinction curves (colors) as compared with our measurements (thick gray line).  The bottom panel shows the derivative of the slope with respect to $R(V)$ (or $\alpha$ in the case of FM09, with an arbitrary rescaling) for different filter combinations.  A sense for the uncertainty in our measurements is given by the thin gray lines, which show our measurements on 10 different temperature subsamples of the APOGEE data.  The top panel shows that several extinction curves provide acceptable descriptions of the reddening in the optical (M14, F99, FM04, FM09), but only FM09 also provides an acceptable description in the infrared.

We note that Figure~\ref{fig:redrvcurves} highlights the disagreement between extinction curves as much as possible.  The slopes of the reddening vectors in Figure~\ref{fig:redrvcurves} are related to second derivatives of the extinction curve.  An alternative would be to plot $E(\lambda - r)/E(g-r)$.  In such plots, all of the various extinction curves have similar behavior, since all curves predict that $E(g-\mathrm{infrared})/E(g-r)$ increase with $R(V)$, and all curves look similar in the mean.  So the mixed agreement of Figure~\ref{fig:redrvcurves} is possible despite the fact that ultimately all of these curves share broad similarities in their description of extinction.

For completeness, we also consider the \citet{ODonnell:1994} extinction curve.  This curve is identical to the CCM extinction curve in the infrared, but refines CCM in the optical.  We find that the curve fits our data about as poorly as CCM and M14 do ($\chi^2 = 611$ on 7 degrees of freedom), due to the mismatch in the infrared.  Restricting to the PS1 bands improves the fit only somewhat, obtaining $\chi^2 = 143$ on 2 degrees of freedom.

\subsection{Linking our Extinction Curve with $R(V)$}
\label{subsec:rvlink}

\begin{figure}[htb]
\dfplot{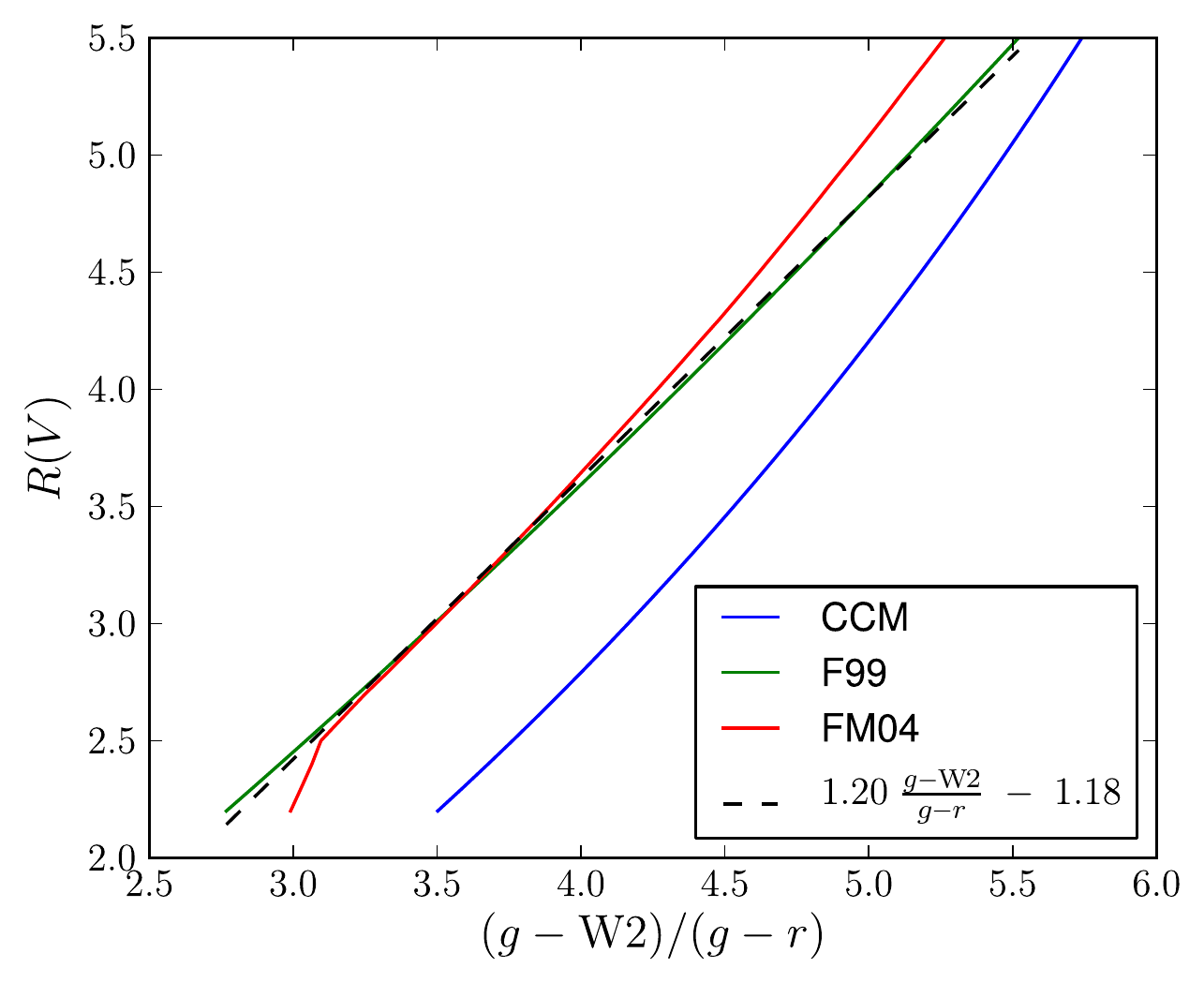}
\caption[$R(V)$ proxy fit]{
\label{fig:rvf99linfit}
The ratio $E(\gps-\mathrm{W2})/E(\gps-\rps)$ for a variety of extinction curves at $A(V) = 2$, as a function of $R(V)$.  The ratio is very close to linear in $R(V)$, motivating us to adopt the proxy $R(V) \approx 1.2 E(\gps-\mathrm{W2})/E(\gps-\rps) - 1.18$ as a simple proxy for $R(V)$, based on a fit to the \citet{Fitzpatrick:1999} extinction curve.
}
\end{figure}

It is useful to link our description of the extinction curve with typical descriptions in terms of $R(V) = A(V)/E(B-V)$.  We cannot directly measure $R(V)$ because it depends on the gray component of the extinction.  However, since $B-V$ is similar to $\gps-\rps$ in that they are both optical colors covering similar wavelengths, and since $A(V)$ is similar to $E(\gps-\mathrm{W2})$, in that $A(\mathrm{W2}) \ll A(\gps)$, we are motivated to look for a linear relationship between $R(V)$ and $E(\gps-\mathrm{W2})/E(\gps-\rps)$.  Figure~\ref{fig:rvf99linfit} shows our calculations for this quantity as a function of $R(V)$ for a variety of extinction curves at $A(V) = 2$.  In all cases, the relationship is not far from linear, though the F99 and FM04 predictions for $E(\gps-\mathrm{W2})/E(\gps-\rps)$ differ from the CCM predictions by as much as 0.5.  All extinction curves predict similar slopes, so we fit a line to the F99 predictions to obtain
\begin{equation}
\label{eq:rvproxy}
\RpV = 1.2 E(\gps-\mathrm{W2})/E(\gps-\rps) - 1.18 \, ,
\end{equation}
and use this as a proxy for $R(V)$.  In terms of our description of the reddening vector in Equation~\ref{eq:redvector}, this roughly corresponds to $R(V) = 3.3 + 9.1 x$.

\subsection{$R(V)$ Variation in the Galaxy}

We have identified a single $R(V)$-like parameter which describes the shape of the extinction curve in the optical and UV.  In this subsection, we make measurements of this parameter for every star in our sample, and use them to study the extent to which $R(V)$ varies in the Galaxy.

\begin{figure}[htb]
\dfplot{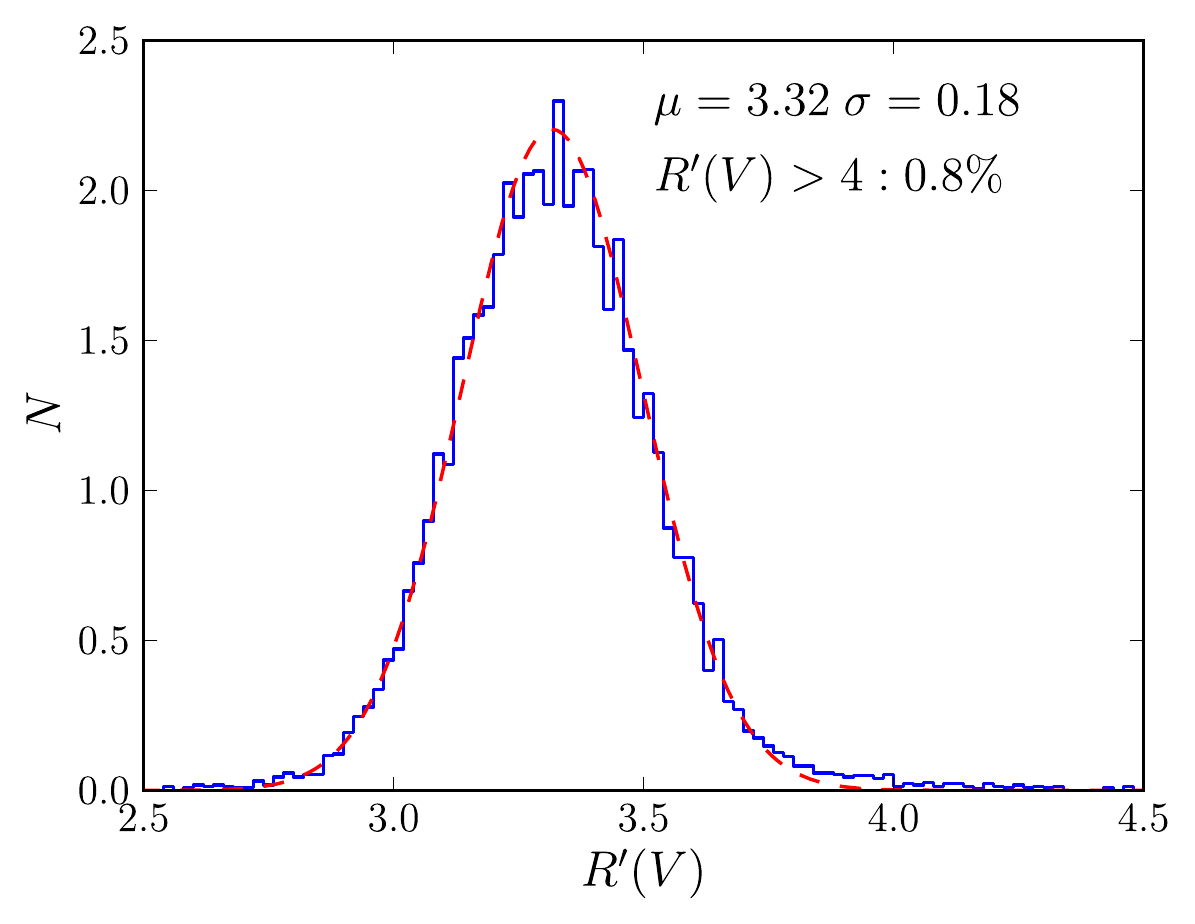}
\caption[$R(V)$ histogram]{
\label{fig:rvhist}
Distribution of $\RpV \approx R(V)$ according to the equation $\RpV = 1.2 E(\gps-\mathrm{W2})/E(\gps-\rps) - 1.18$ (Equation~\ref{eq:rvproxy}).  The width of the distribution of \RpV\ among these stars is remarkably small (0.18).  Moreover, notably absent is a significant tail in this distribution toward large \RpV: only 0.8\% of the sample has $\RpV > 4$, in comparison with 9.5\% in the sample of \citet{Fitzpatrick:2007}.
}
\end{figure}

We use the $R(V)$ proxy \RpV\ of Equation~\ref{eq:rvproxy} to determine the typical extent to which $R(V)$ varies in the Galactic plane.  Figure~\ref{fig:rvhist} shows the distribution of \RpV\ for our sample, for stars with \gps\ photometric uncertainty of less than $0.1$~mag, $\Ep > 0.5$, and \rps\ and $\mathrm{W2}$ photometry.
The distribution is well described by a Gaussian with a mean of 3.32 and a standard deviation of 0.18.  This is a somewhat tighter distribution of $R(V)$ than found by \citet{Fitzpatrick:2007} ($\sigma = 0.27$), and the same size as inferred by \citet{Schlafly:2010}.  For $\Ep > 0.5$, the typical uncertainty in \RpV\ is less than 0.1; the observed scatter is dominated by the intrinsic width of the \RpV\ distribution.  Moreover, the tail to large \RpV\ is much less pronounced in Figure~\ref{fig:rvhist} than in the work of \citet{Fitzpatrick:2007}, possibly owing to the very different populations of stars probed in the two works (O stars versus background giants).  We find, for instance, that only 0.8\% of our stars have $\RpV > 4$, in comparison with 9.5\% in the work of \citet{Fitzpatrick:2007}.  Moreover, many of these stars are simply poor fits, presumably owing to spurious photometry from one of the surveys; the true fraction of high $R(V)$ sight lines is presumably still smaller.  Our results should much better describe the variation in $R(V)$ expected along a typical line of sight.

\begin{figure}[htb]
\dfplot{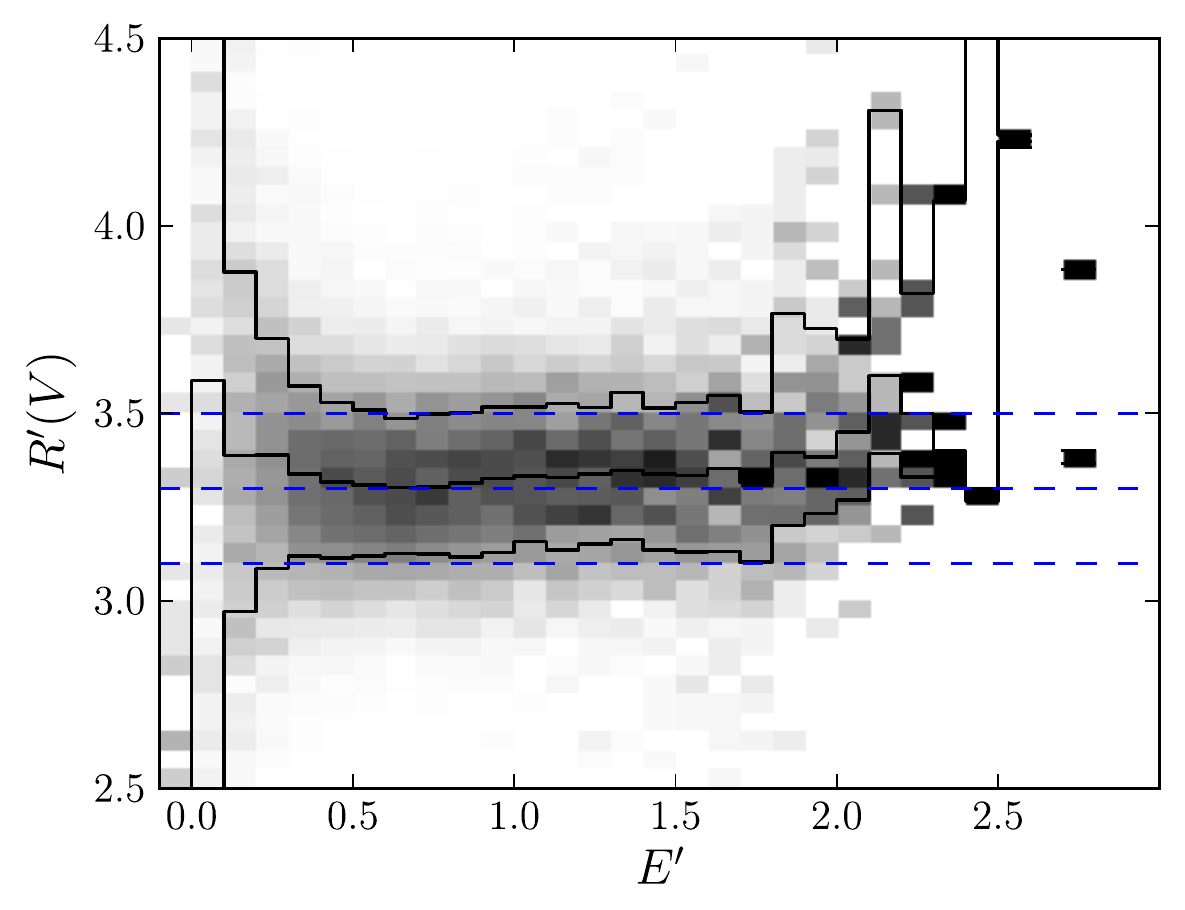}
\caption[$R(V)$ versus $E(B-V)$]{
\label{fig:rvvsebv}
Distribution of \RpV\ versus \Ep, which are roughly equivalent to $R(V)$ and $E(B-V)$.  The grayscale shows the density of points, and the solid lines show the 16th, 50th, and 84th percentiles of the distribution at each \Ep.  Below $\Ep = 0.5$ the uncertainty begins to increase rapidly, but outside this region \RpV\ is remarkably independent of \Ep, described by a simple Gaussian with a standard deviation of 0.2.  There is a slight tendency at the highest \Ep\ where we have \gps\ measurements ($E(B-V) = 2.5$) for \RpV\ to be slightly higher than average (by less than 0.1); this may however be due to a systematic bias, in that there is less $g$ band extinction at higher \RpV\ and fixed \Ep.
}
\end{figure}

Dust properties are expected to change in different environments: a dust grain in a diffuse, atomic cloud is subject to a very different radiation field and collision rate than a dust grain in a dense molecular cloud.  For example, \citet{Whittet:1988} found signatures of water-ice in infrared spectra of stars in Taurus, along sight lines with $E(B-V) \gtrsim 1$, but sight lines with lower $E(B-V)$ were free of ice.  The work of \citet{Ysard:2013} finds signs of grain agglomeration at similar dust column densities.  The extinction curve might then be predicted to change from $\Ep < 1$ to $\Ep > 1$ mag.  We consider this possibility in Figure~\ref{fig:rvvsebv}.

Figure~\ref{fig:rvvsebv} shows that we find that the distribution of \RpV\ is remarkably independent of \Ep.  When $\Ep < 0.5$ mag, the uncertainty in \RpV\ increases significantly, though to as low as $\Ep = 0.2$ mag we find no significant change in \RpV.  At the highest \Ep, there is a slight tendency for \RpV\ to increase, though the amount is small (0.1) and high \RpV\ stars are easier to observe in \gps, since \gps\ band extinction decreases with \RpV\ at fixed \Ep.  We conclude that there is no trend in \RpV\ with \Ep\ for $\Ep < 2$~mag.  There is some suggestion of a trend for $\Ep > 2$~mag, but we are not confident of its significance because our \gps\ band photometry is insufficiently deep.

We interpret Figure~\ref{fig:rvvsebv} as indicating that there is little change in dust extinction curve properties at $\Ep = 1$~mag, despite the formation of ice mantles found by \citet{Whittet:1988}.  However, we note that \Ep\ is only a rough proxy for dust volume density, as high \Ep\ can either indicate individual dense clouds or a number of diffuse clouds along the line of sight.  Despite this, the majority of our sight lines are in the outer galaxy where often much of the total dust column is found in a single cloud, so we expect \Ep\ to be an acceptable proxy for dust density there.  This ambiguity will be resolved by the APOGEE Reddening Survey, which specifically targets giants in the background of dense regions of local molecular clouds.

\begin{figure*}[htb]
\dfplot{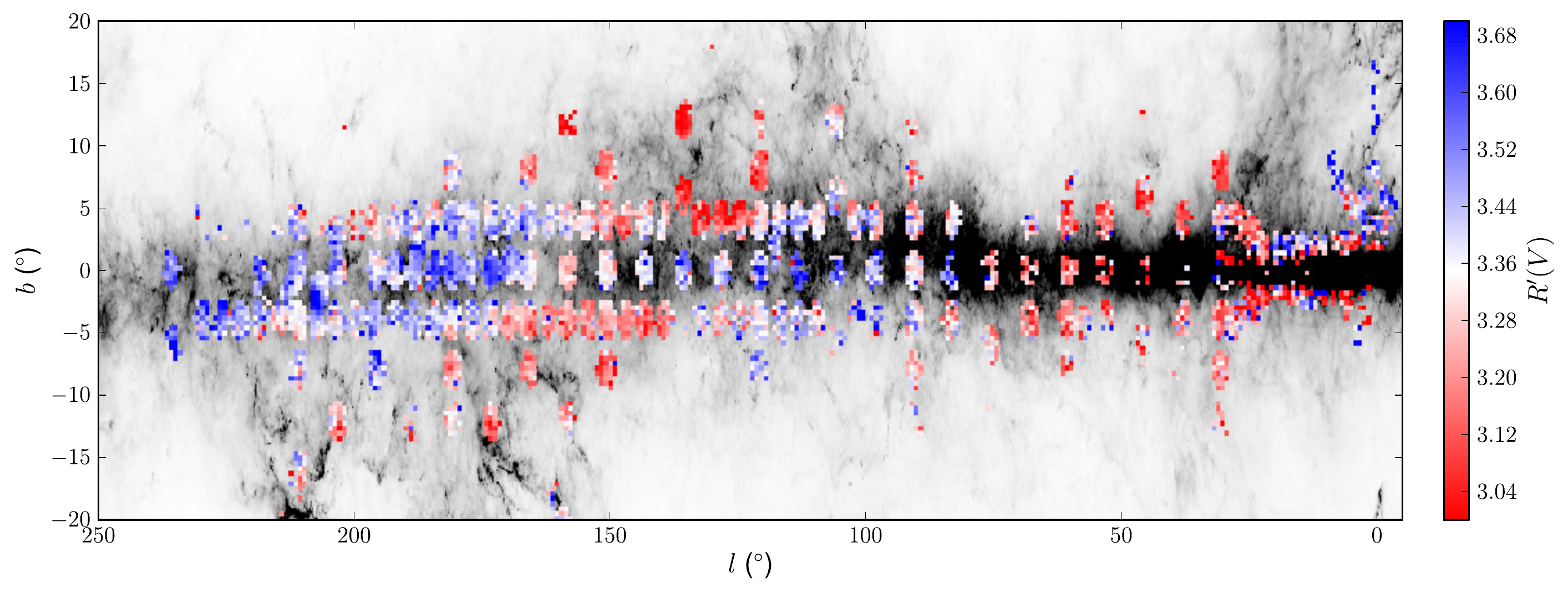}
\caption[$R(V)$ map]{
\label{fig:rvmap}
\RpV\ to APOGEE targets, from Equation~\ref{eq:rvproxy}, for stars with $\Ep > 0.3$~mag.  A map of dust optical depth from \citet{Planck:2014} is provided in the background for context, and ranges from 0--2.5~mag $E(B-V)$.   At $\Ep > 1$, the uncertainty in \RpV\ is typically significantly less than 0.1: the signal is significantly larger than the noise in this map.  Coherent trends in \RpV\ are apparent.  In particular, the Rosette Nebula appears as a region of relatively high $\RpV \approx 4$ at $(l, b) = (206\degree, -2\degree)$, though virtually no other features in the map are easily named.  There is a large band of low \RpV\ dust extending from $(170\degree, -10\degree)$ to $(130\degree, 10\degree)$, and possibly beyond these regions into Orion and Cepheus.  Likewise the dust at $l=50\degree$ has systematically lower \RpV\ than dust at $l=100\degree$.  The rich morphology of the map is poorly correlated with the dust optical depth map and with known ISM structures.
}
\end{figure*}

We can also map the variation in the dust extinction curve over the sky.  Figure~\ref{fig:rvmap} shows the spatial distribution of \RpV\ for the APOGEE targets with $\Ep > 0.3$ mag.  Large, coherent trends in \RpV\ are readily detected.  We detect regions with \RpV\ as low as 2.9 and as high as 3.9.  The most obviously detected cloud with atypical \RpV\ in the APOGEE footprint is the Rosette Nebula, at $(l, b) = (206\degree, -2\degree)$, which is found to have an \RpV\ of about 4.  The extinction curve in the Rosette Nebula was formerly studied by \citet{Fernandes:2012}; they found that most of the stars in the open cluster NGC 2244 follow essentially an $R(V) = 3.1$ extinction curve, though one sight line has $R(V) > 4$.

Except for the Rosette Nebula, however, there are few known structures that appear in the \RpV\ map shown in Figure~\ref{fig:rvmap}.  For example, the APOGEE pointing centered at roughly $(l, b) = (180\degree, 0\degree)$ clearly shows two large clouds at the edges of the field in the $E(B-V)$ map of Figure~\ref{fig:ebvmap}, though the \RpV\ map is featureless.  Likewise for clouds in fields centered at $(l,b) = (140\degree, 0\degree)$ and $(90\degree, 5\degree)$.  Meanwhile some of the most striking features of the \RpV\ map show no features in $E(B-V)$: for instance, the extended region of low \RpV\ centered at $(l, b) = (130\degree, 5\degree)$.

Remarkably, most of the variation in \RpV\ evident in Figure~\ref{fig:rvmap} occurs on scales much larger than an individual molecular cloud.  This is in tension with the traditional picture of $R(V)$ variation as stemming foremost from grain growth in molecular clouds and destruction in feedback in these clouds.  We see only mild evidence for increasing \RpV\ in dense regions of the California Molecular cloud in the APOGEE pointing centered at $(l, b) = (165\degree, -7.5\degree)$, for instance.  Significantly larger variations are apparent over broad regions in Galactic longitude; for instance, $130\degree < l < 170\degree$ has \RpV\ approximately $0.3$--$0.4$ lower than regions of both higher and lower $l$.  This region seems to be correlated with directions in which a majority of the dust column lies within 500~pc in the maps of \citet{Green:2015}, and may also be associated with the edge of the local bubble, but we defer a full characterization of this structure to later work.

We are also in a position to compare the extinction curve in the bulge with the extinction curve more generally in the Galactic plane; just outside of the very inner Galaxy, the extinction looks no different from typical variations within the Galactic disk.  For the innermost Galaxy ($|b| < 2\degree$, $|l| < 20\degree$), most stars are no longer detected in the \gps\ band and we can no longer compute \RpV.  We note, however, that in the inner Galaxy our fits have higher $\chi^2$ than typical elsewhere (\textsection\ref{subsec:fitquality}), so we cannot rule out the possibility that the dust extinction curve there is significantly different from elsewhere in the Galaxy in some way other than $R(V)$.

\subsection{Dust Emission and Extinction Compared}
\label{subsec:rvbeta}

We find measurable variation in the shape of the dust extinction curve.  Likewise, the dust spectral energy distribution (SED) varies significantly.  This variation is often parameterized by the spectral index $\beta$ of the dust emissivity \citep[e.g.,][]{Planck:2014}, or alternatively by the relative amounts of different types of dust with different optical properties \citep[e.g.,][]{Finkbeiner:1999, Meisner:2015}.  Given that both the emission and extinction from dust are ultimately controlled by optical properties of dust grains, it is interesting to compare the variations in these two quantities.

\begin{figure}[htb]
\dfplot{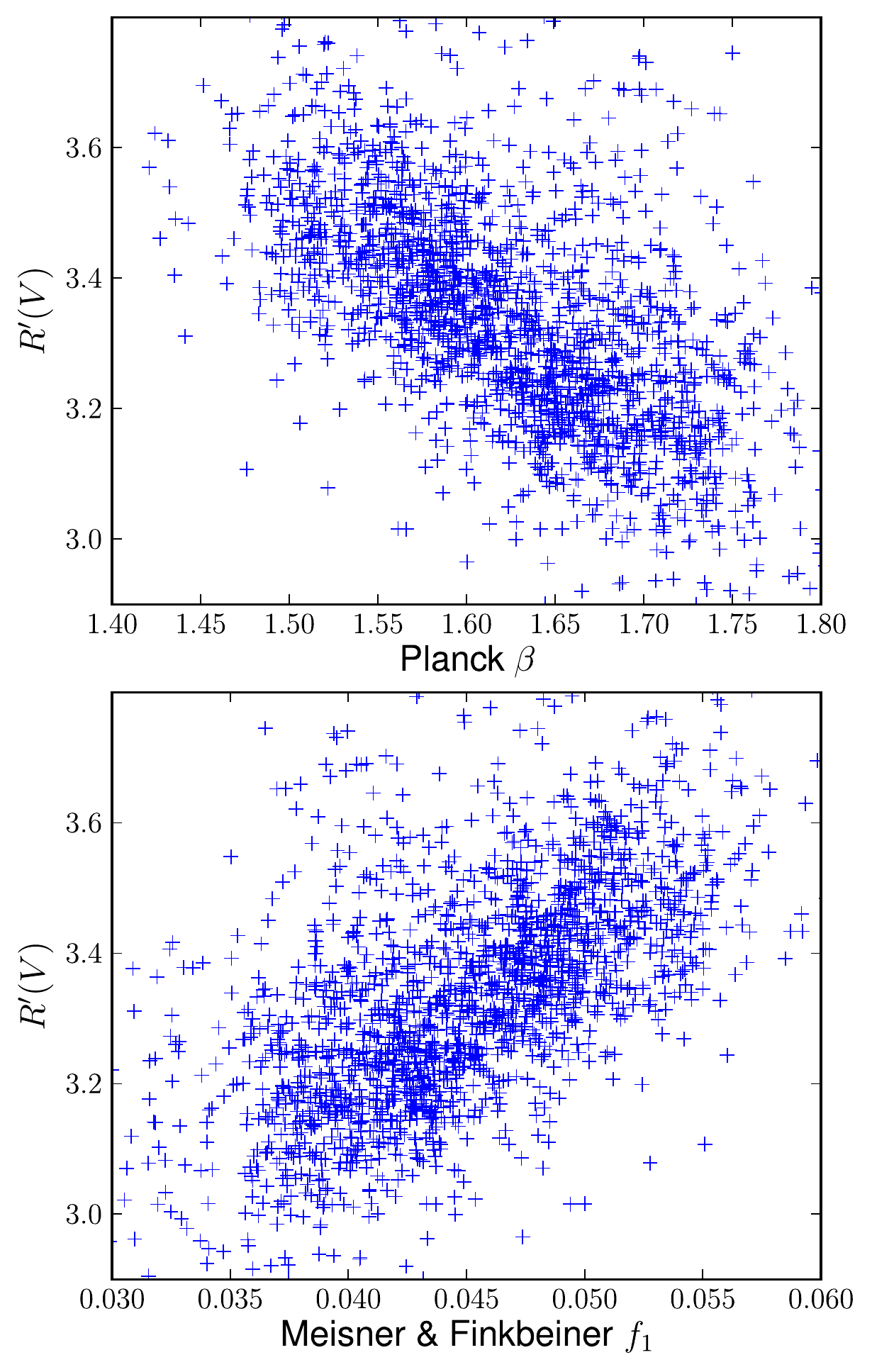}
\caption[$R(V)$ vs. $\beta$]{
\label{fig:betarvscatter}
\RpV\ versus \citet{Planck:2014} $\beta$ and \citet{Meisner:2015} $f_1$, averaged on 1\degree\ scales.  We find a strong correlation between the two quantities, suggesting that variations in the optical-infrared extinction curve and far-infrared SED have a related origin in dust physics.
}
\end{figure}

We show in Figure~\ref{fig:betarvscatter} our measured \RpV\ for stars against the \citet{Planck:2014} $\beta$ and \citet{Meisner:2015} $f_1$ measurements along the same lines of sight, averaged on 1\degree\ scales.  There is a clear relationship: increasing $\beta$ or decreasing $f_1$ corresponds to reducing \RpV.  We note that the \RpV\ measurements and the far-infrared SEDs are statistically completely independent---the first is measured from ground based optical and infrared photometry and spectroscopy of stars, while the second is based on far-infrared measurements of dust emission from space.  It is therefore clear that both are tracing real variations in the properties of the interstellar medium of the Milky Way.  

The discovery that $\beta$ and $R(V)$ are strongly negatively correlated suggests that conditions that lead to steep far-infrared emission spectra also lead to steep optical and infrared extinction curves.  Future models of dust physics will need to accommodate this observational constraint.

The fact that $R(V)$ correlates well with both $\beta$ \emph{and} $f_1$, and not just one or the other, is expected.  The work of \citet{Meisner:2015} models the dust SED as the sum of two modified blackbodies which have different emissivity spectral indices $\beta$.  The second component has $\beta = 2.82$, much larger than the first component, which has $\beta = 1.63$.  Increasing $f_1$ then corresponds to less high $\beta$ dust, leading to a lower effective $\beta$.  This gives rise to a strong correlation between the \citet{Planck:2014} $\beta$ values and the \citet{Meisner:2015} $f_1$ values.  Indeed, we find that $f_1 \approx 0.158 - 0.069 \beta$ over the APOGEE sightlines.  Ultimately the two parameterizations are tracking the largely same variability in the dust SED.

The ideal comparison between dust emission spectra and extinction curves would be performed in high latitude molecular clouds where the stars used for tracing $R(V)$ are behind the entire dust column and where the dust emission is dominated by a single cloud.  The majority of the APOGEE targets, however, are at low Galactic latitudes where both of these conditions are violated.  At these latitudes, even relatively simple measurements like far-infrared optical depth are problematic \citep[e.g.,][]{Schlafly:2014b}.  Ideal sight lines for comparison would also contain no significant CO, which complicates the SED modeling, and would be at high ecliptic latitudes, where zodiacal light is a small contributor to the total far-infrared emission.  However, in Figure~\ref{fig:betarvscatter} we simply include all sight lines.  We therefore expect that the underlying relationship between \RpV\ and $\beta$ or $f_1$ may be significantly stronger, though we currently lack adequate coverage of high latitude clouds in APOGEE to confirm this hypothesis.  The correlation we observe persists regardless of the cuts on CO emission or ecliptic latitude we impose.

\subsection{The ``Gray'' Component of the Extinction}
\label{subsec:gray}

Our analysis is only sensitive to the colors of stars.  Any component of the extinction curve that uniformly extinguishes light across the optical and infrared is undetectable in this analysis.  This insensitivity stems from our ignorance of the distance to any of the sources we measure.  This is a serious limitation, because many practical applications of the extinction curve require knowledge of, for instance, $A(J)/A(H)$, which we are unable to measure.

Many measurements from the literature (e.g., CCM) determine $A(V)$ from reddenings using a fixed extinction curve in the infrared.  This is essentially the solution we adopt in the Appendix, but we are hesitant to employ this procedure, since we seek to measure the variation of the extinction curve.

The simplest solution to this problem would be to adopt measurements of the gray component of the extinction curve from studies of globular clusters, the Galactic bulge \citep[e.g.,][]{Stutz:1999, Nataf:2013, Nataf:2015}, or external galaxies, like the SMC, LMC, or Andromeda \citep[e.g.,][]{MaizApellaniz:2014, DeMarchi:2015}, where the distances to all of the stars are known.  Because the variations in \RpV\ we measure are presumably linked to variations in the gray component of the extinction curve, we need measurements of the gray component over a wide range of \RpV, and preferably not limited to a single star forming region.  This may be possible in Andromeda with the PHAT survey \citep{Dalcanton:2012}, but mapping the gray component of extinction throughout the Milky Way will have to wait for parallaxes from the Gaia mission \citep{Perryman:2001}.

\subsection{The Intrinsic Colors of Giants Compared with Synthetic Models}

Our model produces estimates of the intrinsic colors of giant stars as a function of their temperature and metallicity.  Because faint, metal-rich giants are almost always located far away and in the disk, they are typically significantly reddened.  Our colors may therefore be some of the best empirical estimates of the intrinsic colors of these stars in the combined PS1, 2MASS, and WISE bands, useful for informing models.  We note however two limitations: first, we are ultimately tied to the $\yps-K$ color from the MARCS synthetic stellar grid, and there are hints of problems of around a few hundredths with those models (\textsection\ref{subsec:intcolors}).  Second, we are projecting observed reddened colors back to intrinsic colors across a typical $A(V) \approx 2$~mag linearly along the reddening vector; we should be considering the full non-linear effect of reddening on magnitudes here, which may make a difference of a couple hundredths in the optical.

\begin{figure*}[htb]
\dfplot{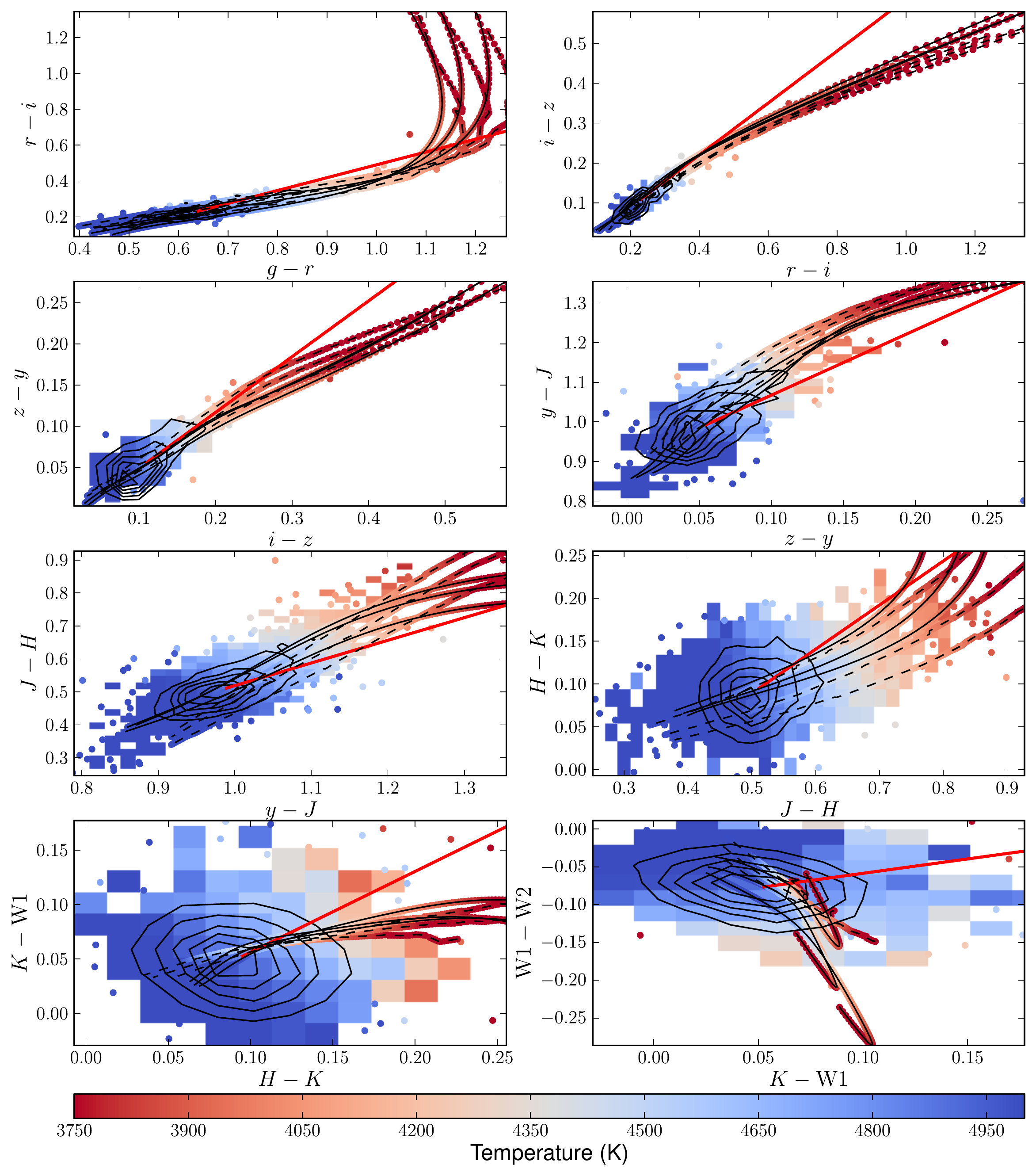}
\caption[Observed, Model, and Synthetic Intrinsic Colors]{
\label{fig:obs-model-syncol}
Observed, SFD-dereddened colors of APOGEE stars, compared with the model intrinsic colors we determine and synthetic colors from the MARCS spectral grid.  Points are colored by their corresponding temperature, and in dense regions we have replaced the points with a colored bin giving the average temperature of all points in that bin.  The contours show the density of points.  The three solid lines show the model colors for metallicities of $-0.75$, $-0.25$, and $0.25$, while the three dashed lines show the synthetic MARCS colors for the same metallicities.  The color along each line shows the temperature of the model.  In general, there is good agreement between the observed dereddened colors (points), the model intrinsic colors (solid lines), and the synthetic colors (dashed lines).  The biggest differences appear for the coldest stars, where the synthetic models are expected to have difficulties due to the formation of molecules.
}
\end{figure*}

Figure~\ref{fig:obs-model-syncol} shows the intrinsic colors we derived, as compared with observed, SFD-dereddened colors and with synthetic colors from the MARCS grid, for stars with $E(B-V)_\mathrm{SFD} < 0.2$.  Observed colors are shown by circles, colored by their temperature.  The model colors of \textsection\ref{subsec:initialfit} are shown by the solid black lines; the three lines correspond to $\mathrm{[Fe/H]} = (-0.75, -0.25, 0.25)$ and the color along the line corresponds to temperature.  The model colors match the observed colors well (up to photometric noise in the observed colors, most obvious at long wavelengths).  Synthetic MARCS colors are shown with dashed lines.  Generally, the synthetic and model colors are in close agreement, though differences of up to several hundredths are present.  Offsets of similar size were also found by \citet{Schlafly:2011} when comparing observed colors of stars with MARCS models, and may be partially due to errors in the APOGEE temperature and metallicity scales.

The largest differences between the model colors and the synthetic colors occur for the coolest stars, $T < 4000~\mathrm{K}$.  For these stars, the onset of molecule formation makes their synthetic colors especially uncertain, so color differences are expected.

In general, our model colors provide a better match to the observed, SFD-dereddened colors than the synthetic colors.  However, very few of the coolest $T < 4000~\mathrm{K}$ stars are present in the $E(B-V)_\mathrm{SFD} < 0.2$ sample where we believe the SFD-dereddening to be adequate, making it hard to assess which of the two is actually more accurate in this region.  That said, we expect our technique to perform nearly as well for cold stars as warm stars, subject only to the accuracy of our fit in Figure~\ref{fig:intyk}, which itself depends on an accurate MARCS $\yps-K$ synthetic color.

\section{Conclusion}
\label{sec:conclusion}

We present sensitive measurements of optical-infrared reddenings to 37,000 stars in the Galactic disk, enabled by APOGEE spectroscopy and PS1, 2MASS, and WISE photometry.  The typical star has a reddening of $0.65$ mag $E(B-V)$, and the stars probe the dust over much of the disk within a few kiloparsecs, making for a uniquely powerful set of reddening measurements.

We use these reddening measurements to determine the shape of the extinction curve in the Milky Way and its variation.  We draw the following conclusions:
\begin{itemize}
\item We make new measurements of the mean extinction curve in the optical through infrared.  Agreement is good with past measurements and the extinction curve of FM09, but other extinction curves (CCM, F99, FM04, M14) provide poor matches to the full optical-infrared extinction curve.
\item We find that the shape of the extinction curve in the optical and infrared can be well characterized by a single parameter, for instance, $R(V)$.  The curvature of the extinction curve increases with decreasing $R(V)$ throughout the optical and infrared, with smaller variation in the near-infrared than in the optical.
\item The shape of the extinction curve is surprisingly uniform, with $\sigma(R(V)) \approx 0.18$, and fewer than 1\% of sightlines having $R(V) > 4$.
\item The variation in $R(V)$ that does exist is uncorrelated with column density for $E(B-V) < 2$.
\item The variations in $R(V)$ we observe are spatially coherent on large scales ($>30\degree$), suggesting that most of the observed variation in $R(V)$ is driven by processes that act on large scales.  In particular, the lack of correlation with column density suggests that the variation is tracing much more than grain growth in dense molecular clouds.
\item Finally, we discover a previously unknown, strong correlation between the thermal dust SED and the shape of the extinction curve.
\end{itemize}

Our work leaves at least two important questions unanswered.  First, we are unable to measure any gray component of the extinction, or the variation in this gray component with the variation we observe in the extinction curve.  Addressing this question is crucial to providing measurements of the full extinction curve $A(\lambda)$, rather than the reddening curve $E(\lambda - \lambda_0)$.  Up to now, extinctions have been much more challenging to measure than reddenings, but the upcoming release of data from Gaia will resolve this long-standing problem.

Second, we have discovered large, coherent variations in the shape of the dust extinction curve, and a relative absence of expected small-scale variations in dense regions.  These variations are strongly negatively correlated with maps of the FIR spectral index of the dust SED.  We are aware of no adequate theoretical framework for understanding the large-scale variations.  In future work, we plan to better characterize ISM structures leading to these signals, in the hopes of providing observational clues to the source of the variations.  Recent improvements in 3D dust mapping \citep{Schlafly:2014, Green:2015} will facilitate this effort.

The forthcoming APOGEE Reddening Survey, part of APOGEE-II, will also help to address this issue.  The survey targets bright red giants in the background of the densest parts of several nearby molecular clouds: Orion, Perseus, Taurus, and Monoceros-R2.  APOGEE temperatures, metallicities, and gravities for these stars combined with PS1, 2MASS, and WISE photometry will allow us to study the shape of the extinction curve in the densest parts of local molecular clouds, indicating to what extent the extinction curve varies in dense regions.

ES acknowledges support for this work provided by NASA through Hubble Fellowship grant HST-HF2-51367.001-A awarded by the Space Telescope Science Institute, which is operated by the Association of Universities for Research in Astronomy, Inc., for NASA, under contract NAS 5-26555, and funding from Sonderforschungsbereich SFB 881 ``The Milky Way System'' (subproject A3) of the German Research Foundation (DFG).  DF acknowledges support of NASA grant NNX10AD69G.  GMG and DPF are partially supported by NSF grant AST-1312891.  The work of JK was supported by the Deutsche Forschungsgemeinschaft priority program 1573 (``Physics of the Interstellar Medium'').  N.F.M. gratefully acknowledges the CNRS for support through PICS project PICS06183.

Funding for the SDSS and SDSS-II has been provided by the Alfred P. Sloan Foundation, the Participating Institutions, the National Science Foundation, the U.S. Department of Energy, the National Aeronautics and Space Administration, the Japanese Monbukagakusho, the Max Planck Society, and the Higher Education Funding Council for England. The SDSS Web Site is http://www.sdss.org/.

The SDSS is managed by the Astrophysical Research Consortium for the Participating Institutions. The Participating Institutions are the American Museum of Natural History, Astrophysical Institute Potsdam, University of Basel, University of Cambridge, Case Western Reserve University, University of Chicago, Drexel University, Fermilab, the Institute for Advanced Study, the Japan Participation Group, Johns Hopkins University, the Joint Institute for Nuclear Astrophysics, the Kavli Institute for Particle Astrophysics and Cosmology, the Korean Scientist Group, the Chinese Academy of Sciences (LAMOST), Los Alamos National Laboratory, the Max-Planck-Institute for Astronomy (MPIA), the Max-Planck-Institute for Astrophysics (MPA), New Mexico State University, Ohio State University, University of Pittsburgh, University of Portsmouth, Princeton University, the United States Naval Observatory, and the University of Washington.

This publication makes use of data products from the Wide-field Infrared Survey Explorer, which is a joint project of the University of California, Los Angeles, and the Jet Propulsion Laboratory/California Institute of Technology, funded by the National Aeronautics and Space Administration.

This publication makes use of data products from the Two Micron All Sky Survey, which is a joint project of the University of Massachusetts and the Infrared Processing and Analysis Center/California Institute of Technology, funded by the National Aeronautics and Space Administration and the National Science Foundation.

The Pan-STARRS1 Surveys (PS1) have been made possible through contributions of the Institute for Astronomy, the University of Hawaii, the Pan-STARRS Project Office, the Max-Planck Society and its participating institutes, the Max Planck Institute for Astronomy, Heidelberg and the Max Planck Institute for Extraterrestrial Physics, Garching, The Johns Hopkins University, Durham University, the University of Edinburgh, Queen's University Belfast, the Harvard-Smithsonian Center for Astrophysics, the Las Cumbres Observatory Global Telescope Network Incorporated, the National Central University of Taiwan, the Space Telescope Science Institute, the National Aeronautics and Space Administration under Grant No. NNX08AR22G issued through the Planetary Science Division of the NASA Science Mission Directorate, the National Science Foundation under Grant No. AST-1238877, the University of Maryland, and Eotvos Lorand University (ELTE).

\appendix
\label{app:extcurve}

We have made new measurements of the extinction curve in the optical through infrared with an unprecedentedly sensitive sample of reddening targets, and find substantial disagreement with existing extinction curves.  Accordingly, we wish to provide a new extinction curve consistent with our data.

Unfortunately, our measurements are lacking in three important ways.  First, we measure the extinction curve only in broad photometric bands, and are largely insensitive to the shape of the curve within those bands.  Second, we are insensitive to any gray component of the extinction curve---but important quantities like $A(\lambda_1)/A(\lambda_2)$ require knowledge of the gray component.  Third, we are only able to measure the extinction curve from the \gps\ to $\mathrm{W2}$ bands.

To address the first of these problems, we simply interpolate between the broad photometric bands we measure with a cubic spline, acknowledging that we are ignorant of the detailed shape.  To address the second, we use the measurement $A(H)/A(K) = 1.55$ from \citet{Indebetouw:2005} to fix the gray component of the extinction curve.  This latter procedure is problematic if there is significant variation in the infrared extinction curve, but at least that is the least variable part of the extinction curve that we observe.  Despite this limitation, the extinction curves we derive by this technique look reasonable, and this procedure provides a simple way to fix the gray component observationally.  To address the third problem, we can only caution the reader that the extinction curve is unreliable outside 5000--45000 \AA, and that at the edges of this range the slope is relatively uncertain.  At the edges of the spline we set the third derivative of the spline to zero, an arbitrary choice.

We construct the extinction curve by first determining the monochromatic wavelengths to which the broad band measurements of Table~\ref{tab:rab} apply.  In analogy with the definition of isophotal wavelengths, we define isoextinction wavelengths $\lambda^e_b$
\begin{equation}
A(\lambda^e_b)/A_0 = \frac{d}{dA_0} \left( -2.5 \log \int d\lambda F(\lambda) 10^{-A(\lambda)/2.5} T_b(\lambda) \right) \, ,
\end{equation}
where $A(\lambda)$ is the extinction at the wavelength $\lambda$, $A_0$ is the extinction at some reference wavelength, $F(\lambda)$ is the flux from a star, and $T_b(\lambda)$ is the total system throughput in the band $b$.  Roughly, the isoextinction wavelengths are the wavelengths at which the monochromatic extinction equals the rate of change of extinction, in magnitudes, in the broad photometric band $b$.

\begin{figure}[htb]
\begin{center}\includegraphics[width=0.5\textwidth]{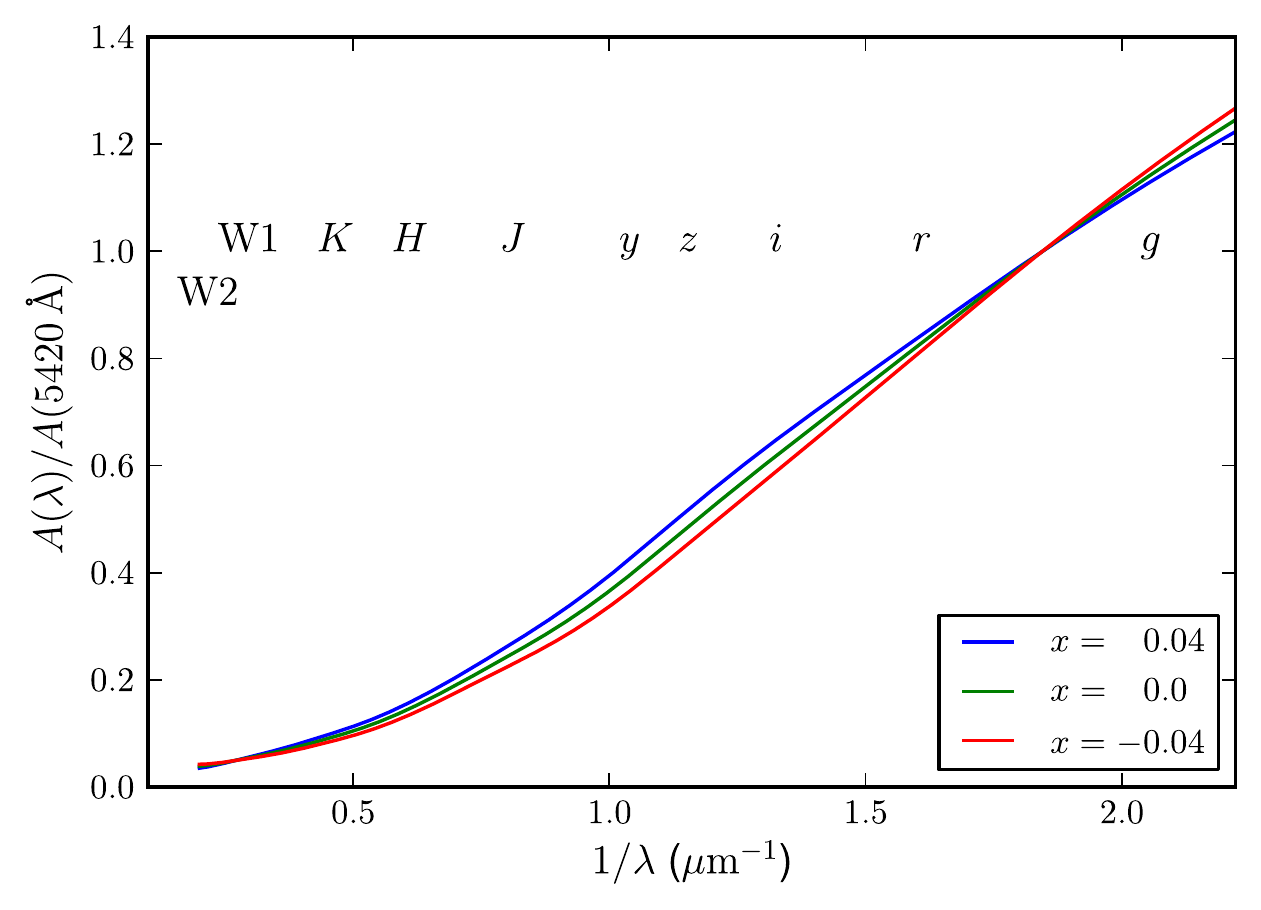}\end{center}
\caption[Extinction Curve]{
\label{fig:extcurves16}
Extinction curves for $x=0.04, 0.0$ and $-0.04$, according to our APOGEE measurements.  Curves are normalized at 5420~\AA, and have gray components fixed according to $A(H)/A(K) = 1.55$ from \citet{Indebetouw:2005}.
}
\end{figure}

Given the isoextinction wavelengths, a smooth, monochromatic extinction curve reproducing our measurements is given by a cubic spline connecting our measured $dA_b/dA_0$, as given in Table~\ref{tab:rab}.  The full procedure is then:
\begin{itemize}
  \item Produce an extinction vector as $\vec{A} = \R + x \dRdx$, with \R\ and $\dRdx$ taken from Table~\ref{tab:rab}.
  \item Fix the gray component, by sending $\vec{A} \rightarrow \vec{A} + C$, so $A(H)/A(K) = 1.55$ \citep{Indebetouw:2005}.
  \item The extinction curve $A(\lambda)$ is given by a cubic spline passing through $\vec{A}$ at the wavelengths $\lambda^e_b$ (Table~\ref{tab:rab}), additionally imposing that the third derivative of the spline is zero at the boundaries.
\end{itemize}
Code implementing the above procedure is available at our web site\footnote{\texttt{http://faun.rc.fas.harvard.edu/eschlafly/apored/extcurve\_s16.py}}.  Figure~\ref{fig:extcurves16} shows three example extinction curves, for $x=0.04, 0.0,$ and $-0.04$, corresponding roughly to $R(V) = 3.6, 3.3,$ and $3.0$.

This procedure is somewhat circular.  The isoextinction wavelengths depend on the shape of the extinction curve $A(\lambda)$, but we use these wavelengths to determine the extinction curve.  We resolve this circularity by initially setting $A(\lambda)$ to be the extinction curve of F09, solving for the isoextinction wavelengths, and using those wavelengths to construct our own curve.  This process is then iterated with the new extinction curve until the isoextinction wavelengths have converged.

In detail, the isoextinction wavelengths depend on the amount of extinction and the spectrum of the source.  We find the isoextinction wavelengths corresponding to a 4500~K star with $\mathrm{[Fe/H]} = 0$ and $\log g = 2.5$ at a reddening $g-r = 0.65$, roughly typical of our sample.  The isoextinction wavelengths also depend on the shape of the extinction curve.  However, for changes of $R(V)$ of 0.3, the wavelengths change by less than one part in a thousand, so we neglect this variation, and always use the isoextinction wavelengths corresponding to our mean extinction curve.

One can compute $R(V) = A(V)/E(B-V)$ directly from our extinction curve.  For $x = 0$, the curve gives $R(V) = 3.6$, significantly larger than traditionally associated with the diffuse ISM.  However, the central wavelength of the Landolt $B$ band is roughly 500~nm blueward of the central wavelength of the PS1 $g$ band, our bluest band, so computing $R(V)$ via this method requires extrapolating our extinction curve.  If instead the extinction curve of F99 and this work are spliced together, we obtain an $R(V)$ lower by a few tenths; direct computation of $R(V)$ is sensitive to how the extinction curve is extrapolated.

\bibliography{2dmap}

\end{document}